\documentclass[%
superscriptaddress,
onecolumn,
nofootinbib,
 amsmath,amssymb,
 aps,
]{revtex4-2}
\usepackage{todonotes}%
\usepackage{braket}
\usepackage{graphicx}
\usepackage{dcolumn}
\usepackage{bm}
\usepackage{xcolor} 
\graphicspath{{./}}
\usepackage{amsmath,amssymb,amsfonts}
\usepackage{booktabs}
\DeclareMathOperator*{\argmin}{argmin}

\begin{document}

\preprint{APS/123-QED}

\title{Non-Perturbative Geometric Framework for Qubit Gates under Always-On Couplings}

\author{Junkai Zeng}
\email{zengjunkai@iqasz.cn}
 \affiliation{International Quantum Academy, Shenzhen, 518048, China}
 \altaffiliation{These authors contributed equally to this work.}
 \author{Lin Chen}
 \affiliation{{Southern University of Science and Technology, Shenzhen 518055, China}}
 \altaffiliation{These authors contributed equally to this work.}
\author{Xiu-Hao Deng}
\affiliation{International Quantum Academy, Shenzhen, 518048, China}
\affiliation{Shenzhen Branch, Hefei National Laboratory, Shenzhen, 518048, China}
\email{dengxiuhao@iqasz.cn}

\date{\today}

\begin{abstract}
Always-on couplings in qubit arrays enable multi-qubit entanglement but also generate crosstalk. We present a geometric framework for a driven qubit at fixed detuning, in which the SU(2) evolution traces a curve on a sphere whose radius is determined by the detuning and the control pulse is the curve's geodesic curvature, reducing gate design to drawing a curve. Closed loops yield crosstalk-free single-qubit gates, with a zero-area condition for undetuned subspaces, while nonzero-area loops also permit entangling-gate constructions. The framework is non-perturbative in the conditional detuning, within a dispersive block-diagonal approximation valid when the transverse mixing $g/\Delta$ is small, and recovers perturbative dynamically corrected gate methods as its flat-space limit. Noise robustness can be imposed as an additional constraint. We show numerically that in two- and three-qubit Heisenberg chains the single-qubit gates are robust to coupling and frequency noise, and outperform a representative perturbative robust gate by over an order of magnitude where perturbation theory breaks down.
\end{abstract}

\maketitle

\section{Introduction}\label{sec:Introduction}

Quantum computing has advanced rapidly over the past two decades, with proof-of-principle demonstrations of computational advantages across platforms ranging from superconducting circuits to trapped ions and semiconductor spin qubits~\cite{Review2023,Lau2022,kim2023evidence,gonzalez2021scaling,yoneda2018quantum,xue2022quantum,huang2024high}. Practical quantum computation, however, requires advancing beyond the current Noisy Intermediate-Scale Quantum (NISQ) era~\cite{Preskill2018quantumcomputingin,preskill2023quantum}. A central challenge is to reduce quantum operation errors below the quantum error correction threshold~\cite{knill2000theory,gottesman2000fault,gottesman2013fault}. Addressing this challenge requires sophisticated qubit control strategies for mitigating unwanted interactions both within the quantum system and between systems and their surrounding environment.

In many quantum computing architectures, direct and always-on qubit couplings are often used to simplify engineering and minimize decoherence channels, particularly in solid-state systems such as superconducting qubits~\cite{Strauch2003,Yamamoto2003,Deng2017,wei2024native} and spin qubits in quantum dots~\cite{Loss1998,Burkard1999,Yoshinaga2021}. However, this approach introduces a fundamental challenge. Always-on inter-qubit interactions create unavoidable crosstalk~\cite{Heinz2022,Seedhouse2021}, which can be a key issue in the scaling up of qubit processors. This can be a challenge even for tunable coupler designs as they also struggle to completely eliminate these interactions because of the fundamentally limited off/on ratio. In such always-on architectures, single-qubit gates are no longer the easy case, because they must actively undo the very interaction that two-qubit gates exploit as a resource, making their control problem often as demanding as entangling-gate design. To achieve high-fidelity gates in qubit arrays, it is therefore necessary to eliminate the impact of these unwanted couplings on both the operating and idle qubits during single-qubit operations. Numerous strategies have been proposed to mitigate qubit crosstalk~\cite{Kestner2020Robust,YJHai2024robust,Zhou2023,Watanabe2024,Kanaar2022Cartan,Kestner2023neuralnetworkdesigned,Kestner2023PRB,Kestner2022filterfunctions,hai2025scalable}, yet most rely on computationally intensive numerical optimization. Recent studies~\cite{YJHai2024robust,hai2025geometric,Watanabe2024,hai2025scalable} treat interactions as small perturbations around zero and mitigate crosstalk by analytically designing noise-resistant quantum gates; a representative pulse from this perturbative robust-control family is the principal benchmark in our numerical comparisons. However, quantum systems can operate in regimes of always-on, nonzero couplings that exceed perturbative approximations. Such perturbative constructions become less accurate when the conditional detuning is no longer small relative to the drive. This motivates an analytical framework that describes how couplings impact qubit dynamics without such an expansion, enabling suppression of crosstalk and improvement of gate fidelity through better gate implementation schemes.

In this paper, we introduce a non-perturbative geometric framework for the evolution of a driven qubit at fixed detuning, and use it to design gates in always-on coupled arrays. The central result is a closed-form geometric criterion. A single-qubit gate is immune to the ZZ-induced detuning at arbitrary strength relative to the drive, within the block-diagonal model that holds when the transverse mixing $g/\Delta$ is small, whenever the corresponding SU(2) trajectories trace closed loops on the associated 2-spheres (with an additional net-zero enclosed-area condition when zero-detuning subspaces are present), and the pulse waveform is read off directly as the loop's geodesic curvature. 
This framework is inspired by reverse engineering techniques~\cite{Barnes2013,Barnestopologicalwinding2015}, used for example in the SWIPHT method~\cite{Barnes2013} to mitigate quantum leakage. In particular, a similar spherical representation of driven-qubit evolution at constant detuning was introduced in ~\cite{Barnestopologicalwinding2015}, and its equivalence to a constant-torsion space curve was established in~\cite{ZengJK2019}. Other previous geometric approaches to noise robustness~\cite{ZengJK2019,barnes2022dynamically,YJHai2024robust,buterakos2021geometrical,hai2025geometric,walelign2024dynamically} use Euclidean curves in $\mathbb{R}^3$ whose closure conditions encode perturbative noise cancellation. \cite{buterakos2021geometrical} also considered finite always-on Ising couplings within a multiqubit geometric framework. The geometric method presented in this manuscript differs by working directly with the spherical evolution trajectories to derive explicit geometric conditions for eliminating coupling-induced gate mismatch. The perturbative Euclidean error-curve condition is recovered in the small-detuning limit. 
Single-qubit gates are our primary focus, because the gate must undo the very coupling that entangling gates exploit. The same construction also permits entangling operations through nonzero enclosed solid angle, as illustrated in Appendix~\ref{app:different windings}. We demonstrate the approach numerically through high-fidelity single-qubit $\pi$-gates in two- and three-qubit chains with always-on Heisenberg couplings of strength comparable to the drive amplitude. When robustness constraints are included, the gates maintain high fidelity even in the presence of slow noise in both qubit frequencies and coupling strengths.

\section{Results}\label{sec:Results}

\subsection{System model and crosstalk formulation}\label{sec:Model}

\begin{figure}[h]
\centering
\includegraphics[width=0.9\textwidth]{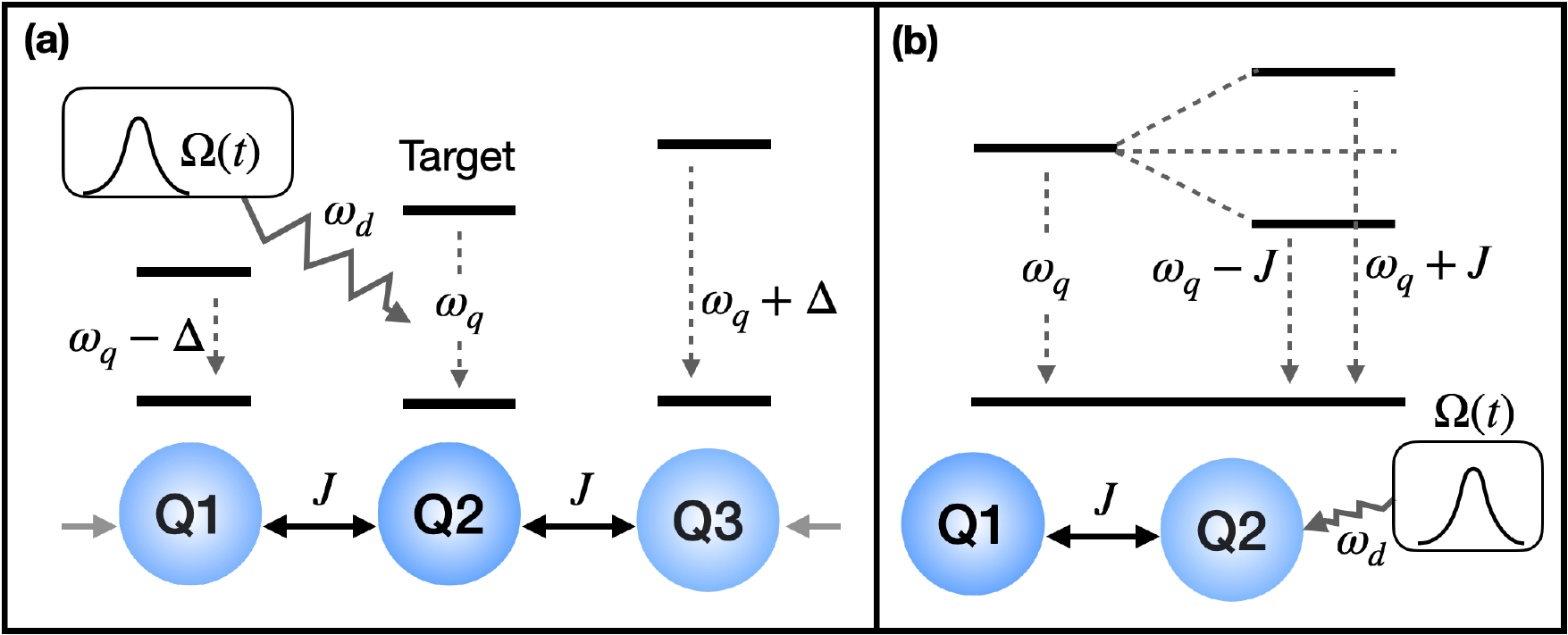}
\caption{\textbf{Schematic of the coupled qubit system.} (a) A three-qubit chain in which the central qubit is driven by a control pulse of frequency $\omega_d$ and is coupled to its neighbors with strength $J$; the qubits have equal frequency separation $\Delta$ from left to right. (b) How the target qubit's frequency splits in a two-qubit system. The target qubit's frequency in (a) similarly splits into three distinct values $\{\omega_q,\omega_q+J,\omega_q-J\}$ depending on the states of the neighbors.}\label{fig:Model}
\end{figure}

We consider multi-qubit systems with always-on couplings.
The couplings in typical quantum architectures, including spin qubits and superconducting transmon qubits, take the form of the combination of $ZZ$-type interactions along with transverse $XX+YY$ terms. As a representative example, consider a one-dimensional qubit chain with identical nearest-neighbor $ZZ$-couplings of strength $J$ and transverse couplings of strength $g$, described by the Hamiltonian:
\begin{equation}
\begin{split}
    H &= \sum_i \frac{\omega_{i}}{2} Z_{i} + \frac{\Omega(t)}{2} \left(\cos(\omega_d t)X_j+\sin(\omega_d t)Y_j\right) \\
      &+ \sum_{i=1}^{N-1} \left(\frac{J}{4} Z_i Z_{i+1}+\frac{g}{4} \left(X_i X_{i+1}+Y_i Y_{i+1}\right)\right)
    \label{eq:heisenberg}
\end{split}
\end{equation}
where $\omega_i$ denotes individual qubit frequencies and $\{X_i, Y_i, Z_i\}$ are the Pauli operators. To implement a single-qubit gate on a target qubit $j$, we apply a control drive with frequency $\omega_d$ and envelope $\Omega(t)$.
We focus on nearest-neighbor interactions, as crosstalk typically decays rapidly with distance in most physical implementations. Two simplified local model systems are relevant in our analysis: two coupled qubits, and one-dimensional three-qubit chains with the central qubit being driven, as schematically depicted in Fig.~\ref{fig:Model}.

Two different types of crosstalk arise from these couplings. The first type, referred to as ZZ-crosstalk, is an energy splitting effect in which the frequency of the target qubit shifts based on the states of neighboring qubits due to $ZZ$-couplings. The second type, referred to as quantum control crosstalk, is the delocalization of the driving field, whereby the control pulse will not only drive the target qubit but also partially impact neighboring qubits. This occurs because the transverse $XX+YY$ coupling causes the Hamiltonian's eigenbasis to deviate from a simple product basis of individual qubits, instead forming `dressed states'. As a result, when the logical qubits are defined by these dressed states, a cross-control effect emerges. The Hamiltonian for the logical qubit in the rotating frame takes the form $\tilde{H}=\bigoplus_{i}\tilde{H}_i+\tilde{V}_{\text{cr}}$, where each $2\times2$ block is:
\begin{equation}\label{eq:qubitblock}
\tilde{H}_i = \frac{1}{2} \left(\beta_i Z + \Omega(t) X \right)
\end{equation}
where $\beta_i$ is the detuning from the driving frequency in that particular subspace, normally on the order of $J$. The number of distinct $\beta_i$ values corresponds to the number of distinct eigenvalues of $\sum_{i\in \text{neighbors}}Z_i$: a qubit with one neighbor (two-qubit configuration) has two $\beta$'s, while one with two neighbors (three-qubit chain) has three, since the ZZ shifts from the two neighbors cancel when they are in opposite states ($|01\rangle$ and $|10\rangle$ thus share the same $\beta = 0$).

The quantum control crosstalk term $\tilde{V}_\text{cr}$ captures the delocalization of the driving field across neighboring qubits:
\begin{equation}\label{eq:controlcrosstalk}
    \tilde{V}_\text{cr}=\sum_{i\in \text{neighbors}} \epsilon_i \Omega(t)Z_{j}\otimes \left(a^\dagger_{i}\exp(-i\tilde{\Delta}_i t) + \text{H.c.}\right)
\end{equation}
where $\epsilon_i\sim g/\Delta_i$ and $\Delta_i=\omega_i-\omega_j$. After time-averaging the fast oscillation at frequency $\tilde{\Delta}_i\sim \Delta_i$, the residual gate error scales as $g\Omega/\Delta_i^2$ and is negligible compared to ZZ-crosstalk when $g\ll \Delta_i$. We therefore focus primarily on ZZ-crosstalk and treat control crosstalk as a perturbative noise term. A detailed derivation for both the two- and three-qubit cases is provided in Appendix~\ref{app:Simplification of the Hamiltonian}.

We formulate the central question as follows: Given a collection of qubit Hamiltonians $\tilde{H}_i$ as in Eq.~(\ref{eq:qubitblock}) with different detunings $\beta_i$, how do we design the control pulse $\Omega(t)$ so that they all evolve to achieve the same desired single-qubit gate operation?

\subsection{2-sphere geometric framework}\label{sec:geometry}

To answer this question we describe the SU(2) evolution of each block as a curve on a 2-sphere whose endpoint depends explicitly on $\beta_i$; designing a single pulse that drives all blocks to the same gate then becomes a geometric problem of making the curves share a common endpoint. In general, SU(2) evolution operators can be parameterized using an Euler-angle-like decomposition, $U(\theta, \chi, \phi; t)=R_X(\theta(t))R_Y(\chi(t))R_X(\phi(t))$, where $R_P(\alpha)=\exp(-i P \alpha/2)$ represents a rotation by angle $\alpha$ about axis $P$. The group manifold of SU(2) is of course $S^3$; what follows is not a statement about that manifold. Rather, for the constrained Hamiltonian considered here the two Euler variables $(\chi,\phi)$ trace a trajectory on a 2-sphere, while the remaining angle $\theta$ specifies the orientation of the moving frame and the final rotation angle.
For a single-qubit Hamiltonian of the form $H(t)=\frac{1}{2}(\Omega_x(t)X+\Omega_y(t)Y+\beta Z)$, applying the inverse Schr\"{o}dinger equation $H(t)dt=i\, dU\, U^\dagger$ and expanding the right-hand side in the Pauli basis gives the differential relations:
\begin{equation}\label{eq:hamchiphi}
\begin{split}
\Omega_x dt &= \cos(\chi)d\phi+d\theta\\
\Omega_y dt &= \left(d\chi, \sin(\chi)d\phi\right)\cdot \left(\cos(\theta), \sin(\theta)\right)\\
\beta dt &=d\chi \sin(\theta) - \sin(\chi)d\phi \cos(\theta)
\end{split}
\end{equation}

We focus on a scenario with a single control field and set $\Omega_y(t)=0$. This condition imposes a constraint on $\theta$, ensuring that the unit vector $\left(\cos(\theta),\sin(\theta)\right)$ is orthogonal to $\left(d\chi, \sin(\chi)d\phi\right)$ at all times. Combined with Eq.~(\ref{eq:hamchiphi}) this yields $|\beta|dt = \left|\left(d\chi, \sin(\chi)d\phi\right)\right|$. The initial condition $U(0)=I$ then fixes $\chi(0)=0$, $\theta(0)=\pi/2$, and $\phi(0)=-\pi/2$.

\begin{figure}[h]
\centering
\includegraphics[width=0.6\textwidth]{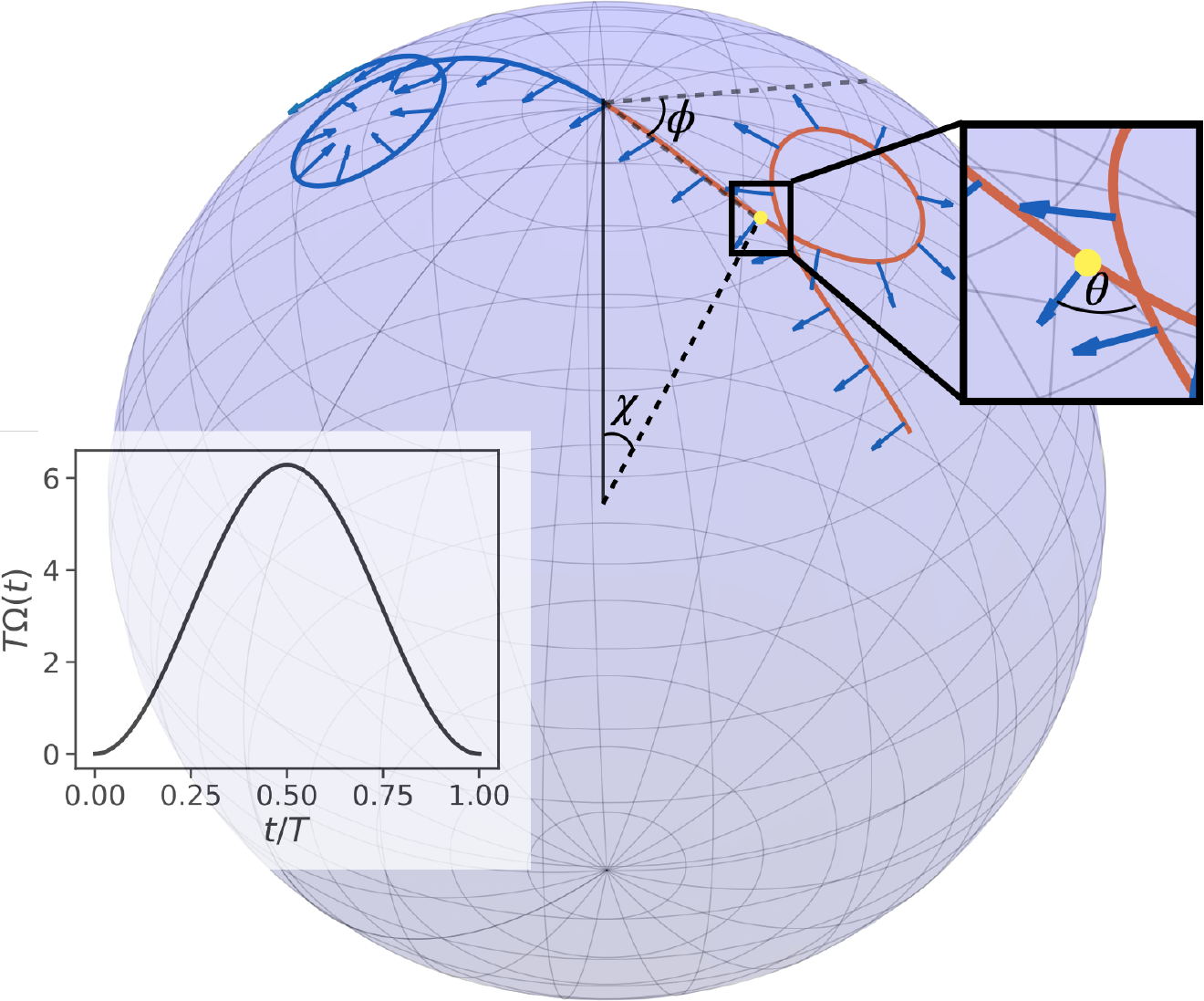}
\caption{\textbf{Geometric representation of the unitary evolution for a $\pi$-gate.} The trajectories on the sphere (blue and orange curves) correspond to Hamiltonian $H=\frac{1}{2}(\pm Z + \Omega(t) X)$ with $\Omega(t)=\frac{\pi}{T}(1-\cos(2\pi t/T))$ (shown in inset), defined by Euler angles $\{\chi, \phi\}$, and exhibit symmetry about the polar axis. The arrows along each curve represent $(\cos(\theta), \sin(\theta))$ and are orthogonal to the tangent vector at all times, enforcing $\Omega_y(t)=0$.}\label{fig:sphericalcurve}
\end{figure}

Geometrically, we can interpret $(\chi(t),\phi(t))$ as spherical coordinates on a 2-sphere (polar and azimuthal angles, respectively). The evolution operator $U(\theta, \chi, \phi;t)$ then traces out a curve $\gamma$ on the surface of a sphere of radius $\frac{1}{|\beta|}$, given by $\vec{\gamma}=\frac{1}{|\beta|}\{\sin(\chi)\cos(\phi),\sin(\chi)\sin(\phi),\cos(\chi)\}$. The sphere radius scales inversely with the magnitude of the detuning because a larger $|\beta|$ accelerates the SU(2) precession, compressing the geometric arc traced per unit time. The arclength of the curve equals the elapsed evolution time, $t(s_0)=\frac{1}{|\beta|}\int_0^{s_0}\sqrt{\sin(\chi(s))^2\phi'(s)^2+\chi'(s)^2}ds$. In this and the following paragraph we take $\beta>0$ for definiteness; the case $\beta<0$ follows by the reflection described below. Fig.~\ref{fig:sphericalcurve} illustrates a trajectory associated with the implementation of a $\pi$-gate using a cosine-shaped pulse, which serves as a standard baseline against which other pulse shapes can be compared.

In terms of the Darboux frame on the sphere, the normal curvature $\kappa_n(t)$ and geodesic curvature $\kappa_g(t)$ provide useful geometric interpretations. A direct computation of $\gamma'$ and $\gamma''$ from $\vec\gamma$ above, combined with Eq.~(\ref{eq:hamchiphi}) under $\Omega_y=0$, yields $\kappa_n(t)=-|\beta|$ and $\kappa_g(t)\equiv |\beta| \vec\gamma \cdot (\vec\gamma' \times \vec\gamma'')=\Omega_x(t)$, where primes denote derivatives with respect to arclength, which by the relation above equals elapsed time.  To see this more explicitly, write $U^\dagger \sigma_a U=\mathbf{e}_a\cdot \mathbf{\sigma}$, where $a=x,y,z$. From Schroedinger equation, we obtain 
$\dot{\mathbf e}_x=-\beta\mathbf e_y$, $\dot{\mathbf e}_y=\beta\mathbf e_x-\Omega_x\mathbf e_z$. For $\beta>0$, choosing $\gamma=-\mathbf{e}_x/\beta$, up to a fixed rotation of the coordinate axes, gives $\ddot{\boldsymbol\gamma}=-\beta^2\boldsymbol\gamma-\Omega_x\mathbf e_z$, whose normal and tangential components give the stated curvature relations. 
The detuning controls bending normal to the sphere; the control pulse controls bending within it. Thus, drawing a particular curve $\gamma(t)$ on the sphere directly specifies, through its geodesic curvature, the control pulse $\Omega_x(t)$ that realizes it. This 2-sphere construction is distinct from the Aharonov--Anandan geometric phase setting~\cite{Berry1984,AharonovAnandan1987}, in which closed loops in projective Hilbert space yield observable phases; here the closed-loop condition instead matches evolution operators across different detunings.

Each $2\times 2$ Hamiltonian block with detuning $\beta_i$ generates its own curve on a sphere of radius $1/|\beta_i|$, all sharing the north pole as the common starting point ($\chi(0)=0$). Suppressing crosstalk amounts to choosing one $\Omega_x(t)$ that simultaneously drives all these curves, drawn on different spheres, to a common endpoint. For two-qubit systems, the detunings may be $\beta_1 = -\beta_2$ if the driving frequency is set midway between the two split energy levels, or include a zero detuning $\beta_1=0$ if the driving pulse resonates with one of the split qubit frequencies. For the three-qubit case, the detunings may be $\beta$, $-\beta$, and 0, respectively.

The two configurations treated here therefore give the symmetric detuning sets $\{\pm\beta\}$ and $\{0,\pm\beta\}$, for which the construction is analytic: the $-\beta$ block closes automatically once the $+\beta$ block does, by the reflection symmetry described below, and the $\beta=0$ block is accounted for by the single scalar condition of Eq.~(\ref{eq:zeroarea}). A general collection of unequal magnitudes $|\beta_i|$ does not inherit this structure, since a curve that closes on one sphere need not close on a sphere of different radius; closure must then be imposed on each radius as a separate constraint within the same optimization. The existence and optimization of such solutions are left for future work.

From Eq.~(\ref{eq:hamchiphi}), imposing $\Omega_y=0$ makes reversing the sign of $\beta$ equivalent to reflecting the curve about the polar axis (i.e., $\chi\rightarrow -\chi$), as shown in Fig.~\ref{fig:sphericalcurve}. When $\beta=0$ (resonant driving), the curve remains at the north pole ($\chi=0$). To achieve the same final unitary operation for different values of $\beta$, including $\beta=0$, we must ensure that all corresponding curves begin and end at the same point (north pole): $\chi(0)=0$ and $\chi(T)=2n\pi$, where integer $n$ counts the number of times the curve wraps around the sphere. For $\beta \neq 0$, this closed-loop curve yields an X-rotation gate $R_X(\Phi)$ with $\Phi=\Delta\theta+\Delta\phi$. However, for $\beta=0$, the evolution is $R_X(\int\Omega_x(t)dt)=R_X(\Delta \theta + \oint_\gamma \cos(\chi) d\phi)$, which differs from $\Phi$ in general. Matching the rotations therefore requires an additional condition on the enclosed solid angle. Using $\Delta\phi=\oint_\gamma d\phi$ for a closed curve, the condition becomes $\oint_\gamma(1-\cos(\chi))d\phi=0$; geometrically, the closed curve must subtend zero net signed solid angle:
\begin{equation}
\label{eq:zeroarea}
    S_\gamma\equiv\frac{1}{2}\oint_\gamma (1-\cos(\chi))d\phi=0.
\end{equation}
Throughout, $S_\gamma$ denotes one half of the signed solid angle subtended by the loop. It is a dimensionless functional of the curve's shape alone, and is distinct from the metric area enclosed on the sphere of radius $1/|\beta|$, which carries an additional factor $1/\beta^2$. We refer to $S_\gamma$ loosely as the enclosed area below, with this convention understood.

More generally, a loop with $S_\gamma\neq0$ leaves the detuned and resonant blocks with unequal rotation angles, $R_X(\Phi)$ and $R_X(\Phi-2S_\gamma)$ respectively. This conditional rotation can also be used to construct entangling gates, as illustrated in Appendix~\ref{app:different windings}.

Finally, the number of twists the curve traces on the sphere affects the rotation angle $\Phi=\Delta\theta+\Delta\phi$ of the gate $R_X(\Phi)$. Any closed curve starting and ending at the north pole must satisfy $\Delta\theta=k \pi$, where the integer $k$ counts how many net loops form along the curve. $\Delta\phi$ measures the total winding around the polar axis. In practice, for curves that cross the south pole ($\chi(T)>0$), it is preferable to forgo such loops so that $\Delta\theta=0$ and $\Phi=\Delta\phi$. We note one topological subtlety: since $R_Y(2n\pi)=(-1)^n I$, curves with odd $n$ pick up a global $-1$ relative to even $n$; choosing even winding (we use $n=2$ throughout) avoids this complication.

\subsection{Noise robustness via Magnus expansion}\label{sec:robust_gate}

The geometric framework above guarantees the target gate for ideal Hamiltonian parameters, but does not by itself address fluctuations in those parameters. A Magnus expansion shows that noise robustness adds one more integral constraint to the same closed curve~\cite{barnes2022dynamically}.
Consider a noisy Hamiltonian component $\delta H_k(t)$, associated with a particular noise source $k$. The random noise strength $\delta k$ is assumed to be small and unknown, and we take $\delta H_k \propto \delta k$. The first-order Magnus term associated with noise-free dynamics $U_0(t)$ can be written as an integral:
\begin{equation}\label{eq:m1}
       A^{\delta H_k}_1(t) = \int_{0}^{t} d\tau U_0^\dagger(\tau) \delta H_k(\tau) U_0(\tau) =\sum_i a^{\delta H_k}_i(t)P_i
\end{equation}
where $\{P_i\}$ is the multi-qubit Pauli operator basis and $a^{\delta H}_{i}(t) =\frac{1}{d}\text{Tr}\left[ P_{i} A^{\delta H}_{1}(t) \right]$ are the associated coefficients with $d=2^n$ the Hilbert-space dimension. We model noise as $\delta H_J = \delta J \sum_{i\in \text{neighbors}}Z_\text{target}\otimes Z_{i}$ for coupling fluctuations and $\delta H_\omega = \delta \omega Z_\text{target}$ for frequency noise. These noises are treated as quasi-static. We also consider the quantum control crosstalk terms $\delta H_\text{cr}=\tilde{V}_{\text{cr}}$ from Eq.~(\ref{eq:controlcrosstalk}) as an additional noise source with small strength $\epsilon_i$.

To achieve first-order noise robustness, we minimize the weighted $L^2$-norm:
\begin{equation}
    |C_\text{robust}|^2 =\sum_k c_k|\partial_{\delta k} A^{\delta H_k}_1(T)|^2
\end{equation}
where $T$ is the gate time, $\delta k \in \{\delta J, \delta \omega, \epsilon_i \}$, and $c_k$ are weighting factors. As shown in Appendix~\ref{app:Analysis of error curve}, a useful consequence of the geometric structure is that the same integral controls both $\delta\omega$ and the dominant part of $\delta J$ susceptibility, so a single optimization simultaneously addresses both noise channels.

\begin{figure}[h]
\centering
\includegraphics[width=\textwidth]{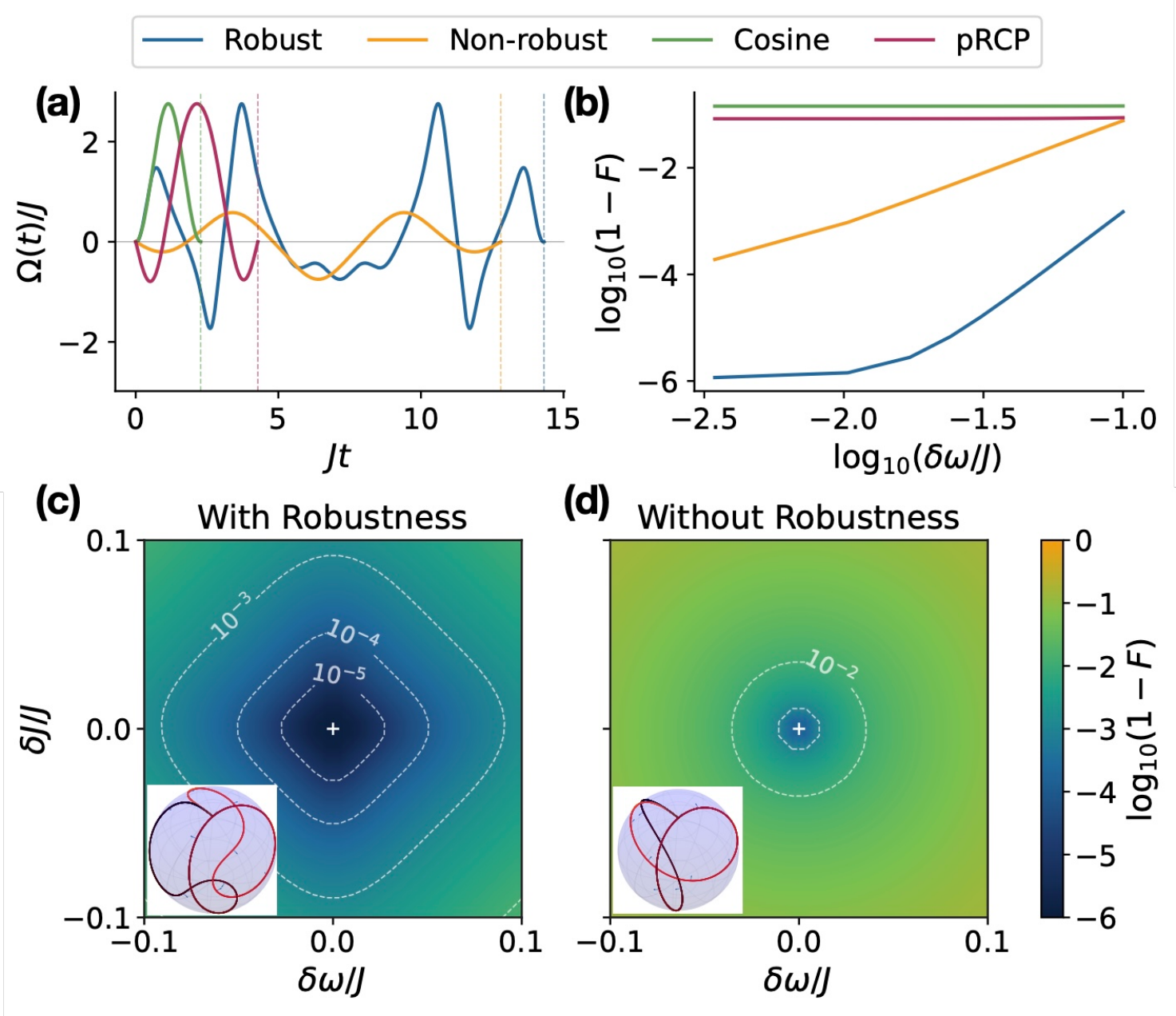}
\caption{\textbf{Single-qubit $R_X(\pi)$ gate in an always-on coupled two-qubit setting.} (a) Driving pulses for four implementations: the robust pulse (minimizing $C_\text{robust}$), the non-robust pulse (enforcing $C_\text{target}=0$ only), a cosine baseline and a pRCP, all sharing the same maximum amplitude. (b) Infidelity $1-F$ as a function of qubit-frequency noise $\delta\omega/J$ at $\delta J = 0$. (c) and (d) Log-infidelity maps for the robust and non-robust scenarios, respectively, under simultaneous noise in the qubit frequency $\delta \omega$ and coupling strength $\delta J$. The insets in (c) and (d) show the associated 2-sphere trajectories (color-coded from red at the start to black at the end). All quantities are in units of the coupling strength $J$, with $J/\Delta=1/20$.}\label{fig:PiPulse_2}
\end{figure}

\begin{figure}[h]
\centering
\includegraphics[width=\textwidth]{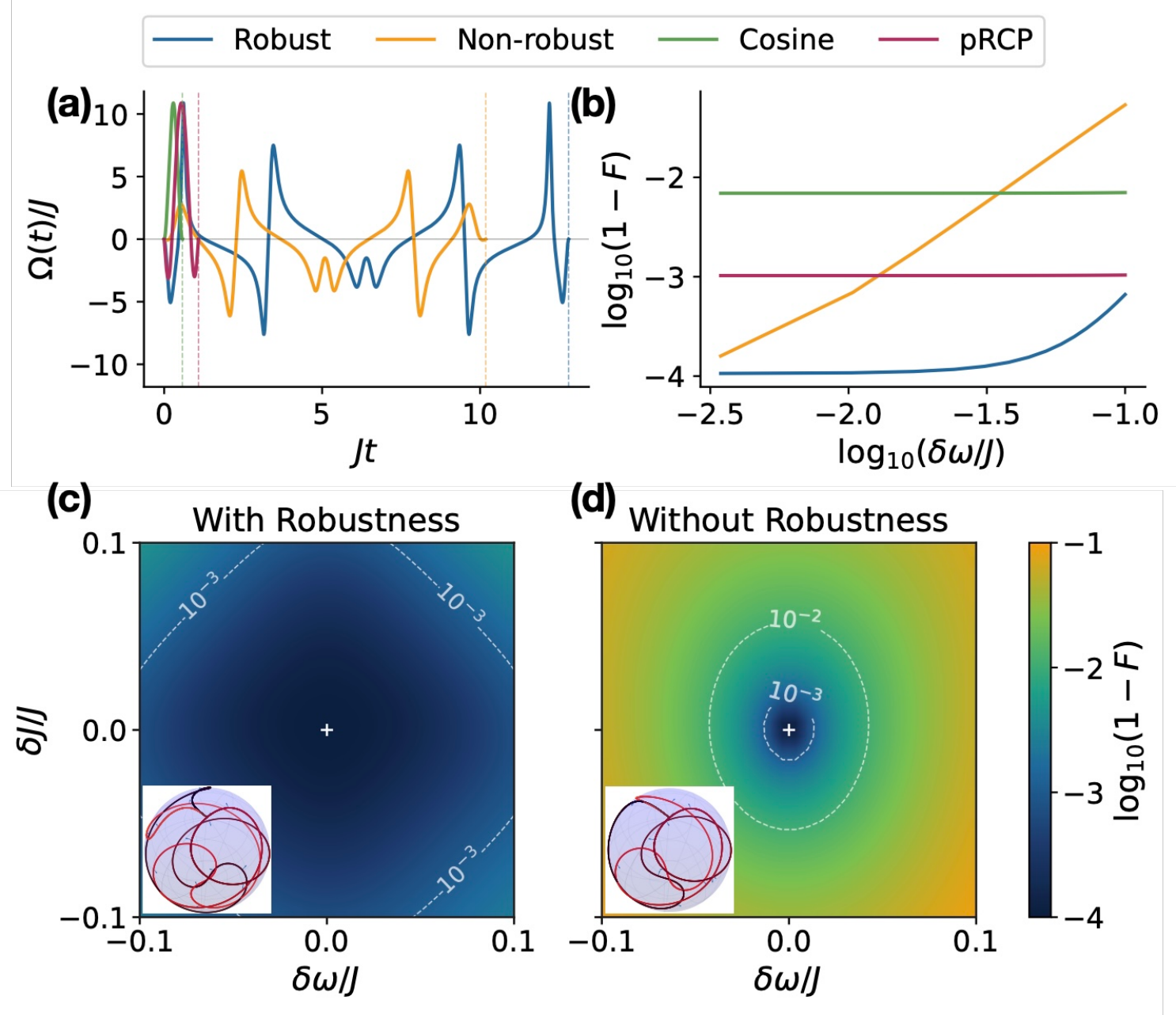}
\caption{\textbf{Single-qubit $R_X(\pi)$ gate in a three-qubit chain with always-on couplings.} (a) Pulse waveforms for the robust pulse, the non-robust pulse, the cosine baseline, and the pRCP; the cosine and pRCP pulses share the same maximum amplitude as the robust pulse. (b) Log-log infidelity versus qubit-frequency noise strength $\delta\omega/J$ at $\delta J = 0$. (c) and (d) Log-infidelity maps for the robust and non-robust implementations as functions of $\delta J$ and $\delta \omega$. All quantities are in units of the coupling strength $J$, with $J/\Delta=1/20$.}\label{fig:PiPulse_3}
\end{figure}

\subsection{Numerical demonstrations}\label{sec:results}

We perform numerical simulations for systems with Heisenberg interaction as given in Eq.~(\ref{eq:heisenberg}) with $g=J$ and $J/\Delta=1/20$. This regime is naturally realized in silicon spin qubits with isotropic exchange coupling, for which $J/h\sim 1$--$10$~MHz and detunings $\Delta/h\sim 20$--$200$~MHz between neighboring qubits are typical~\cite{Loss1998,Burkard1999,Yoshinaga2021,yoneda2018quantum,xue2022quantum,huang2024high}; for such parameters our robust $\pi$-gates complete in $\sim 100$~ns at the upper end of this coupling range and $\sim 1\,\mu$s at the lower end, against reported $T_2^\ast \sim 10$--$100\,\mu$s in isotopically purified $^{28}$Si. We note that at the lower end this is a percent-level fraction of $T_2^\ast$, which is not negligible at the infidelity levels reported below; the simulations here treat quasi-static parameter noise rather than irreversible decoherence, and a quantitative decoherence-limited fidelity would require the dynamical noise spectrum as input. The same $J/\Delta$ range also describes fixed-frequency capacitively-coupled transmons, although their coupling structure is $XX+YY$-dominated rather than fully Heisenberg. The results for Case 1, the implementation of a single-qubit $R_X(\pi)$ gate in a two-qubit coupled system, are shown in Fig.~\ref{fig:PiPulse_2}. We compare four pulse waveforms: a robust pulse obtained by minimizing $C_\text{robust}$; a non-robust pulse that only enforces the target gate; a basic cosine pulse that neglects the residual coupling but matches the maximum amplitude of the robust pulse (used as a baseline); and a perturbative robust control pulse (pRCP) constructed following Ref.~\cite{hai2025geometric}, with the same maximum amplitude, that is designed to be first-order robust against weak transverse noise. The explicit pRCP pulse shape used in our benchmark, together with a brief description of its construction and its connection to the small-$\beta$ limit of the present framework, is given in Appendix~\ref{app:pRCP}. As shown in Figs.~\ref{fig:PiPulse_2}(a) and (b), our approach, despite its much longer gate time relative to the baseline and pRCP, achieves substantially better fidelity. The cosine baseline, designed to perform a fast $\pi$-rotation while ignoring the residual coupling, has a gate time about five times shorter than the robust pulse; despite this speed, it plateaus at $1-F > 10^{-2}$ regardless of noise strength. The pRCP also fails to achieve a high-fidelity $\pi$-gate, with only a marginal improvement over the cosine baseline. The residual infidelity in both cases is a systematic error from the unmitigated always-on coupling. Both pulses are constructed by expanding about $\beta=0$, and the relevant expansion parameter is the conditional detuning measured against the drive, $\beta/\Omega_\text{max}$ or equivalently $\beta T$, rather than the dispersive ratio $J/\Delta$ that controls the block-diagonal reduction of Eq.~(\ref{eq:qubitblock}). Here $\beta/\Omega_\text{max}\approx0.36$ and $\beta T\approx4.3$ for the pRCP, so that expansion is not controlled even though $J/\Delta=1/20$ is small.

We also note the resource convention used in these comparisons. Both benchmark pulses are matched to the robust pulse in peak amplitude, not in duration, so the comparison holds peak drive power fixed and lets gate duration float. The comparison therefore characterizes these particular pulse implementations under a fixed peak-amplitude constraint, rather than their optimal performance at equal gate duration. Figs.~\ref{fig:PiPulse_2}(c) and (d) show that incorporating the noise-robustness condition leads to a longer and more complex pulse, but improves gate fidelity when the system is subjected to noise affecting the coupling strength and the qubit frequency. Even at zero quasi-static noise the robust pulse outperforms the non-robust one, since the same constraints that suppress $\delta\omega$ and $\delta J$ susceptibilities also reduce the unmitigated coherent control crosstalk that scales as $g\Omega/\Delta^2$. The robust pulse achieves $1-F \lesssim 10^{-5}$ at zero noise and remains below $10^{-3}$ across the window $|\delta\omega|, |\delta J| \le 0.05 J$, whereas the non-robust pulse degrades to $\sim 10^{-2}$ over the same window.

Case 2, as depicted in Fig.~\ref{fig:PiPulse_3}, involves the implementation of a single-qubit $R_X(\pi)$ gate in a three-qubit chain. We evaluate the same four implementations as in Case 1: robust, non-robust, cosine baseline, and pRCP. As shown in Figs.~\ref{fig:PiPulse_3}(a) and (b), the cosine baseline and pRCP again have gate times more than an order of magnitude shorter than our approach. Two of the four diagonal blocks here have vanishing detuning ($\beta_2 = \beta_3 = 0$) by symmetry of the three-qubit chain, so the cosine and pRCP pulses experience effectively zero crosstalk on half of the logical subspace; combined with their short gate times, this yields lower infidelity than in Case 1, and even surpasses our non-robust pulse at large frequency noise. Both still fall short of the robust-optimized pulse. The non-robust pulse starts near $1-F \sim 10^{-4}$ at low noise but degrades quadratically, while the robust pulse holds $1-F \lesssim 10^{-4}$ across nearly two decades in $\delta\omega/J$. The effect of residual inter-block control crosstalk beyond the dispersive regime is further examined in Appendix~\ref{app:Parameters of Pulse} by varying $J/\Delta$

\section{Discussion and Outlook}\label{sec:conclusion}

We have presented a geometric framework for the evolution of a driven qubit at fixed detuning, in which gate design reduces to drawing curves on a 2-sphere. The framework is non-perturbative in the conditional detuning, while the reduction to independent $2\times2$ blocks retains a dispersive approximation for transverse state mixing; these are distinct expansions, controlled by $\beta/\Omega_\text{max}$ and $g/\Delta$ respectively. Applied to systems with always-on couplings, it yields both crosstalk-free single-qubit gates and also permits entangling-gate constructions. The central result is a geometric construction of crosstalk-free dynamics. Crosstalk immunity is achieved by forming a closed loop on a 2-sphere; when zero-detuning subspaces are present, this loop must additionally subtend zero net solid angle. The pulse waveform is obtained directly from the curve's geodesic curvature. These conditions provide an explicit geometric prescription for gate synthesis without expanding in the coupling-induced conditional detuning. Adding first-order Magnus-expansion constraints brings noise robustness into the same framework. Numerical demonstrations in two- and three-qubit Heisenberg chains show that the resulting pulses achieve infidelities of $1-F \lesssim 10^{-5}$ at zero quasi-static noise and below $10^{-3}$ across $\sim 5\%$ noise windows, outperforming both a cosine baseline and a representative perturbative robust-control pulse in the regime where the coupling approaches the drive amplitude. The benchmark against pRCP is consistent with the expected advantage of the spherical description over Euclidean-geometric methods once the conditional detuning is comparable to the drive scale.

A few limitations are worth noting. The block-diagonal description that underlies the spherical construction assumes that the inter-block control crosstalk $\tilde V_\text{cr}$ remains parametrically smaller than the block-diagonal detuning; for systems with $g \sim \Delta$ this assumption breaks down and a more general treatment is required. The noise model adopted here is quasi-static, capturing slow drifts in qubit frequencies and exchange strengths but not general time-dependent noise spectra, such as $1/f$ noise, that dominate at long times in solid-state platforms; a filter-function~\cite{Green2013,Kestner2022filterfunctions} extension is the natural next step. The robust pulses are typically longer than perturbative alternatives, trading bandwidth and gate time for non-perturbative crosstalk suppression; the optimal balance between pulse duration, amplitude, and robustness has not been systematically explored. The numerical demonstrations here are restricted to linear nearest-neighbor chains; extensions to other coupling graphs and leakage suppression remain to be carried out.

The same geometric construction also permits entangling operations through nonzero enclosed solid angle. As shown in Appendix~\ref{app:different windings}, this yields a CNOT-equivalent gate in the two-qubit model and an XNOR-CNOT-type operation in the three-qubit chain. These results demonstrate the versatility of the construction, while noise-robust entangling gates remain a direction for future work. Generalization to two-dimensional arrays, dipolar coupling, and all-to-all connectivity would broaden the applicability to neutral-atom and trapped-ion platforms, and to analog quantum simulators where always-on Hamiltonians are intrinsic to the model. More broadly, expressing gate and noise-cancellation conditions as geometric properties of curves, such as closure, enclosed area, and winding, may carry over to other control problems beyond the cases treated here.

\section{Methods}\label{sec:methods}

\subsection*{Gate fidelity}
Gate performance is quantified by the average gate fidelity between the
simulated unitary $U$ and the target gate $U_{\mathrm{target}}$,
\begin{equation}
    F = \frac{d + \left|\mathrm{Tr}(U_{\mathrm{target}}^\dagger U)\right|^2}{d(d+1)},
\end{equation}
where $d = 2^n$ is the Hilbert space dimension. The modulus of the trace
makes this expression invariant under a global phase.

\subsection*{Pulse parameterization and optimization}

The control pulse waveform is determined by designing a closed geometric curve on the 2-sphere. We construct a parameterized function $\phi(\chi;\vec{b})$ with $\chi \in [0,2n\pi]$. Fixing $\chi(T)>0$ avoids the formation of simple loops along the curve, allowing trajectories with smaller geodesic curvature and thus smaller, easier-to-implement pulse waveforms. Boundary conditions are applied: $\phi(T)-\phi(0)$ is set to match the desired rotation angle, and $\phi'(0)=\phi'(T)=0$ ensures the pulse starts and ends smoothly at zero amplitude.

We optimize the parameter vector $\vec{b}$ by minimizing the total cost function $\vec{b} = \argmin_{\vec{b}} C_\text{total}[\phi(\chi;\vec{b})]$,
where $C_\text{total}=w_1|C_\text{target}|^2+w_2 |C_\text{robust}|^2$ combines the target gate fidelity and noise robustness objectives with weighting factors $w_1$ and $w_2$. For single-qubit gates within a two-qubit coupled system, where the driving frequency is set midway between the target split qubit frequencies (Case 1), the target evolution condition is inherently met upon satisfying the boundary conditions, resulting in $C_\text{target}=0$ by construction. For single-qubit gates within a three-qubit chain (Case 2), the function must also satisfy Eq.~(\ref{eq:zeroarea}), which takes the form $C_\text{target} = \int_0^{\chi(T)} (1-\cos(\chi))\phi'(\chi)\, d\chi = 2S_\gamma = 0$. In practice, the form of $\phi(\chi)$ can be chosen so that $C_{\text{target}}=0$ is inherently satisfied, allowing optimization to focus entirely on robustness. The optimization uses JAX~\cite{jax2018github} and Optax~\cite{deepmind2020jax} with automatic differentiation; the explicit ansatz and optimized parameters are given in the following subsection. In this work, we set $\chi(T) = 4\pi$; a more general ansatz with $\chi(T) = 2n\pi$, including a modified version of $\phi(\chi;\vec{b})$ for implementing multi-qubit entangling gates, is discussed in Appendix~\ref{app:different windings}.

\subsection*{Pulse ansatz and optimized parameters}

We use the following ansatz for $\phi(\chi;\vec{b})$, defined on $\chi\in[0,4\pi]$:
\begin{equation}\label{eq:ansatz}
\phi(\chi) = a(\chi-6\pi)\chi^2 + f(\chi),
\end{equation}
where $a=\Phi/(32\pi^3)$ is fixed by the target rotation angle $\Phi$ and
\begin{equation}\label{eq:fchi}
    f(\chi) = \sin^3\!\left(\frac{\chi}{2}\right)\!\left(b_1\sin\frac{\chi}{4}+b_2\sin\frac{3\chi}{4}+b_3\cos\frac{\chi}{2}+c\right).
\end{equation}
The zero-area constraint $C_{\text{target}}=0$ is satisfied exactly by setting $b_3=-16a(3+4\pi^2)+\frac{4096(13b_1-33b_2)}{45045\pi}$, reducing the free parameters to $b_1$, $b_2$, and $c$. When $C_{\text{target}}=0$ is not required (two-qubit midpoint driving), the ansatz simplifies to $b_2=b_3=0$. Optimized parameters for the $X(\pi)$ and $X(\pi/2)$ gates are listed in Tables~\ref{tab:pulseparameters} and \ref{tab:pulseparameters2}.

\begin{table}[h]
\caption{Parameters of the $X(\pi)$-gate ($a=\frac{1}{32\pi^2}$).}
\label{tab:pulseparameters}
\centering
\begin{tabular}{lcccc}
\hline\hline
 Setting & $b_{1}$ & $b_{2}$ & $b_3$ & $c$\\
\hline
two-qubit, non-robust & 0 & 0 & 0 & 0 \\
two-qubit, robust & $-5.8674$ & 0 & 0 & $5.4642$ \\
three-qubit, non-robust & $5.71915$ & 0 & 0 & 0\\
three-qubit, robust & $221.6515$ & $-20.9140$ & $101.2265$ & $-136.5514$ \\
\hline\hline
\end{tabular}
\end{table}

\begin{table}[h]
\caption{Parameters of the $X(\pi/2)$-gate ($a=\frac{1}{64\pi^2}$).}
\label{tab:pulseparameters2}
\centering
\begin{tabular}{lcccc}
\hline\hline
 Setting & $b_{1}$ & $b_{2}$ & $b_3$ & $c$\\
\hline
two-qubit, non-robust & 0 & 0 & 0 & 0 \\
two-qubit, robust & $-2.9338$ & 0 & 0 & $4.8111$\\
three-qubit, non-robust & $2.8596$ & 0 & 0 & 0\\
three-qubit, robust & $124.1078$ & $-12.1260$ & $57.2051$ & $-73.0914$ \\
\hline\hline
\end{tabular}
\end{table}

\begin{acknowledgments}
The authors thank Huiqi Xue and Yao Song for stimulating discussions.

This work was supported by the National Natural Science Foundation
of China (Grant No.~12404566), the Key-Area Research and Development
Program of Guangdong Province (Grant No.~2018B030326001), the Science,
Technology and Innovation Commission of Shenzhen Municipality
(JCYJ20170412152620376, KYTDPT20181011104202253), and the Shenzhen
Science and Technology Program (KQTD20200820113010023).
\end{acknowledgments}

\section*{Author contributions}

J.Z. developed the theoretical framework and was the main contributor
to manuscript writing. J.Z. and L.C. carried out numerical simulations
and data analysis. J.Z. and X.-H.D. supervised the project.
All authors contributed to discussions and manuscript preparation.

\section*{Competing interests}

The authors declare no competing interests.

\section*{Data and code availability}

The source code for geometric pulse optimization and numerical
simulation, together with the notebooks used to generate the results
presented in this work, is publicly available at
\url{https://github.com/zengjk/non_perturbative_robust_control}.


\clearpage
\appendix

\appendix
\numberwithin{equation}{section}
\numberwithin{figure}{section}
\numberwithin{table}{section}

\begin{center}
  \large\bfseries Appendices
\end{center}
\vspace{1ex}

\newcommand{\appendixsection}[1]{%
  \refstepcounter{section}%
  \section*{Appendix \thesection.\ #1}%
}

\appendixsection{Block-Diagonal Hamiltonian Derivation}
\label{app:Simplification of the Hamiltonian}

\subsection{Two-Qubit System}
\label{subsec:two qubit case}

We consider a system of two coupled qubits. The Hamiltonian consists of two parts: $H(t)=H_0+H_c(t)$, where
\begin{equation}
\begin{split}
    &H_0 = \frac{1}{2}(\omega_1 Z_1+\omega_2 Z_2)+\frac{1}{4}\left(g_1 (X_1 X_2+Y_1Y_2)+g_2 Z_1Z_2\right)\\
    &H_c(t)=\frac{\Omega(t)}{2}\left(\cos(\omega_d t) X_2 +\sin(\omega_d t) Y_2 \right)
\end{split}
\end{equation}
Here $\omega_i$ are the individual qubit frequencies,
$g_1$ and $g_2$ are the transverse and ZZ coupling strengths, respectively (in the notation of the main text, $g_1\equiv g$ and $g_2\equiv J$). $\Omega(t)$ is the time-dependent driving field amplitude,
with $\omega_d$ being the driving frequency. For Heisenberg coupling, $g_1=g_2=J$.
We define $\Delta = \omega_1-\omega_2$ as the qubit frequency difference.
We transform to a frame where the undriven Hamiltonian is diagonalized so that the eigenbasis defines the logical qubit, through transformation matrix $S$, $H\rightarrow \tilde{H}= SHS^\dagger$:
\begin{equation}\label{eq:S}
    S = \left(
\begin{array}{cccc}
 1 & 0 & 0 & 0 \\
 0 & \cos (\theta ) & -\sin (\theta ) & 0 \\
 0 & \sin (\theta ) & \cos (\theta ) & 0 \\
 0 & 0 & 0 & 1 \\
\end{array}
\right)
\end{equation}

Here $\theta = -\frac{1}{2}\arctan(g_1/\Delta)$. This transforms the Hamiltonian into a block diagonal form,

\begin{equation}
        \tilde{H}=\left(
\begin{array}{cccc}
 \bar{\omega }+\frac{g_2}{4} & \frac{1}{2} e^{-i t \omega_d} \tilde{\Omega }(t) & \frac{1}{2} \tan (\theta ) e^{-i t \omega_d} \tilde{\Omega }(t) & 0 \\
 \frac{1}{2} e^{i t \omega_d} \tilde{\Omega }(t) & \frac{1}{4} \left(2 \Delta  \sec (2 \theta )-g_2\right) & 0 & -\frac{1}{2} \tan (\theta ) e^{-i t \omega_d} \tilde{\Omega }(t) \\
 \frac{1}{2} \tan (\theta ) e^{i t \omega_d} \tilde{\Omega }(t) & 0 & \frac{1}{4} \left(-2 \Delta  \sec (2 \theta )-g_2\right) & \frac{1}{2} e^{-i t \omega_d} \tilde{\Omega }(t) \\
 0 & -\frac{1}{2} \tan (\theta ) e^{i t \omega_d} \tilde{\Omega }(t) & \frac{1}{2} e^{i t \omega_d} \tilde{\Omega }(t) & \frac{1}{4} \left(g_2-4 \bar{\omega }\right).  \\
\end{array}
\right)
\end{equation}

Here $\bar{\omega}$ is the mean qubit frequency $(\omega_1+\omega_2)/2$, and $\tilde{\Omega}(t)=\cos(\theta)\Omega(t)$ is the effective drive amplitude in the dressed-state frame. (The pulse waveforms plotted in the main-text figures are the effective dressed drive $\tilde{\Omega}(t)$; the corresponding physical drive is $\Omega(t)=\tilde\Omega(t)/\cos\theta$, where the dressing factor differs from unity by $O(g^2/\Delta^2)$ and is invisible at the scales shown.) In this transformed frame, the upper-left and lower-right $2\times2$ blocks represent the Hamiltonian of the target qubit when the other qubit is in two different states, with split qubit frequencies $\tilde{\omega}_2 \pm \frac{g_2}{2}$, where $\tilde{\omega}_2=\omega_2+\frac{1}{2}(\Delta-\sqrt{g_1^2+\Delta^2})$ is the qubit frequency shifted due to the transverse coupling $g_1$. The off-diagonal blocks represent the coherent control crosstalk term that drives the other qubit, which itself has a shifted and split frequency $\tilde{\omega}_1\pm \frac{g_2}{2}$, with $\tilde{\omega}_1=\omega_1-\frac{1}{2}(\Delta-\sqrt{g_1^2+\Delta^2})$.

To further simplify, we move into a rotating frame, $H\rightarrow R^\dagger H R-iR' R^\dagger$, with transformation matrix $R = \exp\left[
-i \left(\tilde{\omega}_1 Z_1 + \omega_d Z_2\right) t \right] $. In this rotating frame, the Hamiltonian becomes:
\begin{equation}
\tilde{H}_\text{rot}=\left(\begin{array}{cc}
\tilde{H}_1 & 0 \\ 0 & \tilde{H}_2
\end{array}\right) +\tilde{V}_\text{cr},
\end{equation}
where each $\tilde{H}_i$ is a $2\times 2$ matrix of the form $\tilde{H}_i=\frac{1}{2}\left(\beta_i Z + \tilde{\Omega}(t) X\right)$, and $\tilde{V}_\text{cr}$ represents the coherent control crosstalk:
\begin{equation}\label{eq:crosstalktwo}
\tilde{V}_\text{cr} = \frac{1}{2}\tan(\theta) \tilde{\Omega}(t) \left(\cos(\tilde{\Delta} t)XZ+\sin(\tilde{\Delta} t)YZ\right).
\end{equation}
If the driving frequency $\omega_d$ is chosen to be the midpoint between the split target qubit frequencies, $\omega_d=\tilde{\omega}_2$, then the parameters in the Hamiltonian are $\beta_1=-\beta_2=g_2/2$ and $\tilde{\Delta} = -\sqrt{\Delta^2+g_1^2}$. Alternatively, choosing the driving frequency resonant with one of the split frequencies, for example $\omega_d=\tilde{\omega}_2- \frac{g_2}{2}$, gives $\beta_1=g_2$, $\beta_2=0$, and $\tilde{\Delta} = -\sqrt{\Delta^2+g_1^2} - g_2/2$.

\subsection{Three-qubit Chain}
\label{subsec:three qubit case}

We now consider a three-qubit system forming a 1D chain, with the driven qubit placed in the middle (indexed as the third qubit so that the Hamiltonian has a clean block-diagonal form). The system Hamiltonian is given by:
\begin{equation}
\begin{split}
    H &= H_0 +H_c(t)\\
    H_0&=\frac{\omega_1}{2} Z_1 + \frac{\omega_2}{2} Z_2 + \frac{\omega}{2} Z_3 + V \\
    V &= \frac{1}{4}g_1(X_1X_3+Y_1Y_3+X_2X_3+Y_2Y_3) +\frac{1}{4}g_2(Z_1Z_3+Z_2Z_3)\\
    H_c(t)&=\frac{\Omega(t)}{2} \left[\cos(\omega_d t)X_3 + \sin (\omega_d t)Y_3 \right]
\end{split}
\end{equation}
As in the two-qubit case, frequency difference between neighboring qubits determines the strength of coherent control crosstalk.
For simplicity, we assume that the frequency detuning between the target qubit and its two neighbors is symmetric, $\omega-\omega_1=\omega_2-\omega=\Delta$, although this assumption is not necessary, as we only require the frequency detunings much larger than the coupling strengths. Diagonalizing this Hamiltonian is much more complicated than in the two-qubit case, as it mixes $|001\rangle$, $|010\rangle$, and $|100\rangle$, as well as $|110\rangle$, $|101\rangle$, and $|011\rangle$. To address this, we use a transformation matrix $S(\alpha, \kappa, \gamma)$ such that $\tilde{H}=SHS^\dagger$, where:
\begin{equation}
    \begin{split}
        &S=S_3 S_2 S_1\\
        &S_1(\alpha)=R(\frac{\alpha}{4}, P_1+P_2)\\
        &S_2(\kappa) =R(\frac{\kappa}{4}, P_1-P_2)\\
        &S_3(\gamma)=R(\frac{\gamma}{2}, P_3)\\
        &P_1 = XZY - YZX + ZXY - ZYX \\
        &P_2 = XIY - YIX - IXY + IYX \\
        &P_3 = XYI - YXI
    \end{split}
\end{equation}
Here, each $R(\alpha, P)$ is a unitary matrix defined as $R(\alpha, P)=\exp(i \alpha P)$ for Hermitian operator $P$, and each term in $P$ is a Pauli tensor product, for example $XZY=X_1\otimes Z_2 \otimes Y_3$. The generators $P_1+P_2$ and $P_1-P_2$ couple the two single-excitation triplets $\{|001\rangle,|010\rangle,|100\rangle\}$ and $\{|110\rangle,|101\rangle,|011\rangle\}$, respectively, while $P_3$ corrects the residual mixing between $|010\rangle$ and $|100\rangle$ (and likewise for the other triplet) generated to second order.

Since exact diagonalization is hard, we adopt a perturbative approach by expanding in the small parameter $\lambda=g_1/\Delta$. We take $\alpha= \alpha_1 \lambda + \alpha_2 \lambda^2$, $\kappa = \kappa_1 \lambda + \kappa_2 \lambda^2$, and impose that all off-diagonal elements vanish to $O(\lambda^2)$. This yields
$\alpha_1 = -\kappa_1 = \frac{1}{2}$, $\alpha_2 = \kappa_2 = -\frac{g_2}{4 g_1}$, and $\gamma = \frac{1}{2}\arctan(\frac{\lambda^2}{4+\lambda^2})$. Residual off-diagonal couplings are then of order $\lambda^3$ and are absorbed into the control-crosstalk term below; the diagonal block parameters $\beta_i$ quoted in the next paragraph are exact in $g_2$ but truncated at $O(\lambda^2)$ in $g_1$.

This transformation, while preserving the target qubit frequency and the ZZ couplings, shifts the neighbors' frequencies to $\tilde{\omega}_1 = \omega-\tilde{\Delta}$ and $\tilde{\omega}_2 = \omega+\tilde{\Delta}$, with $\tilde{\Delta}=\Delta(1+\lambda^2/4+\lambda^4/32)$ a perturbative analog of the dressed splitting $\sqrt{\Delta^2+g_1^2}$ encountered in the two-qubit case.

As in the two-qubit case, we now move to a rotating frame defined by the transformation $R=\exp[-i(\tilde{\omega}_1 Z_1 + \tilde{\omega}_2 Z_2 + \omega_d Z_3)t]$, and choose $\omega_d = \omega$; the Hamiltonian then becomes
\begin{equation}
\tilde{H}_\text{rot}=\left(\begin{array}{cccc}
\tilde{H}_1 & & & \\
 & \tilde{H}_2 & & \\
 & & \tilde{H}_3 & \\
 & & & \tilde{H}_4
\end{array}\right) +\tilde{V}_\text{cr},
\end{equation}
with each $\tilde{H}_i=\frac{1}{2}\left(\beta_i Z + \tilde{\Omega}(t) X\right)$, where $\beta_1=-\beta_{4}=g_2$, $\beta_2=\beta_3 =0$, and $\tilde{\Omega}(t)=\Omega(t)(1-\lambda^2/4)$.
The coherent control crosstalk $\tilde{V}_\text{cr}$ is now given by
\begin{equation}\label{eq:crosstalkthree}
\begin{split}
&\tilde{V}_\text{cr} = \Omega(t)\left(\frac{1}{4}(\tilde{V}^{(1)}_1+\tilde{V}^{(1)}_2)\lambda + \frac{g_2}{8g_1}(\tilde{V}^{(2)}_1+\tilde{V}^{(2)}_2+\tilde{V}^{(2)}_{12}) \lambda^2\right)\\
 &\tilde{V}^{(1)}_1 =  \left(\cos(\tilde{\Delta} t)X_1+\sin(\tilde{\Delta} t)Y_1\right)Z_3 \\
 &\tilde{V}^{(1)}_2 =  \left(-\cos(\tilde{\Delta} t)X_2+\sin(\tilde{\Delta} t)Y_2\right)Z_3 \\
 &\tilde{V}^{(2)}_1 =  -\left(\cos(\tilde{\Delta} t)X_1+\sin(\tilde{\Delta} t)Y_1\right)Z_2Z_3  \\
&\tilde{V}^{(2)}_2 =  Z_1\left(-\cos(\tilde{\Delta} t)X_2+\sin(\tilde{\Delta} t)Y_2\right)Z_3 \\
&\tilde{V}^{(2)}_{12} =\left(\cos(2\tilde{\Delta} t)(X_1X_2+Y_1Y_2)+\sin(2\tilde{\Delta} t)(Y_1X_2-X_1Y_2)\right) X_3
\end{split}
\end{equation}

\appendixsection{Noise and First-Order Error Analysis}
\label{app:Analysis of error curve}

In this section, we analyze the first-order error due to noise in coupling strength ($\delta J$) and qubit frequency ($\delta \omega$) using Magnus expansion. For a two-qubit system, the driving pulse can have a frequency either centered between the split target qubit frequencies or resonant with one of the split frequencies. The noise- and quantum crosstalk-free Hamiltonian in the rotating frame is given by $\Omega(t) IX + J ZZ$ or $\Omega(t)IX + J (IZ+ZZ)$, depending on the driving frequency. The corresponding noise Hamiltonians are $\delta H_J = \delta J ZZ$ or $\delta J (IZ+ZZ)$, and $\delta H_\omega = \delta \omega IZ$ for both cases. For simplicity, we omit the scalar coefficients.

\subsection{Driving Frequency Centered Between Split Frequencies}

In this scenario, the three Pauli operators $IX, ZY, ZZ$ form a closed algebra isomorphic to SU(2), allowing an Euler-like decomposition of the evolution operator as $U_0=R_{IX}(\theta)R_{ZY}(\chi)R_{IX}(\phi)$. Consequently, the first-order error contributions from $\delta H_J$ and $\delta H_\omega$, derived from the Magnus expansion as in Eq.~\eqref{eq:m1}, can be expressed as
\begin{equation}\label{eq:errorcase1}
\begin{split}
    A_1^{\delta H_J}(t)&=\delta J\int_0^t d\tau \left(R_{IX}(-\phi)R_{ZY}(-\chi)R_{IX}(-\theta) ZZ\, R_{IX}(\theta)R_{ZY}(\chi)R_{IX}(\phi)\right)\\
    &=\delta J\, ZZ \int_0^t d\tau \left(R_{IX}(\phi)R_{ZY}(\chi)R_{IX}(2\theta)R_{ZY}(\chi)R_{IX}(\phi)\right)\\
    A_1^{\delta H_\omega}(t)&=\delta \omega\int_0^t d\tau \left(R_{IX}(-\phi)R_{ZY}(-\chi)R_{IX}(-\theta) IZ\, R_{IX}(\theta)R_{ZY}(\chi)R_{IX}(\phi)\right)\\
    &=\delta \omega\, IZ \int_0^t d\tau \left(R_{IX}(\phi)R_{ZY}(\chi)R_{IX}(2\theta)R_{ZY}(\chi)R_{IX}(\phi)\right)\\
\end{split}
\end{equation}
(where $\tau$-dependence of $\theta,\chi,\phi$ is implicit). The second equality in each line uses the anticommutation $\{IX,\, ZZ\}=\{IX,\, IZ\}=0$ to pull $ZZ$ (or $IZ$) past the outer $R_{IX}$ factors, which flips the sign of their arguments and combines them with the inner $R_{IX}(\theta)$ into $R_{IX}(2\theta)$.

The two integrals have identical form, so the noise-free dynamics exhibit the same first-order susceptibility to $\delta J$ and $\delta\omega$. Any control scheme optimized to suppress one therefore automatically suppresses the other.

\subsection{Driving Frequency Resonant with a Split Frequency}

In this scenario, the coupling term becomes $IZ+ZZ$, which is no longer a Pauli-like operator. We decompose the evolution operator as $U_0=R_{\frac{IX-ZX}{2}}(\alpha)R_{IX}(\theta)R_{\frac{IY+ZY}{2}}(\chi)R_{IX}(\phi)$, where $R_{\frac{IX-ZX}{2}}(\alpha)$ commutes with the remaining factors and captures the mismatch between the geometric rotation angle $\Delta\theta+\Delta\phi$ and the integrated pulse area $\int_0^t \Omega(\tau)d\tau$. Here $\alpha(t) = \int_0^t (1-\cos(\chi(\tau)))\phi'(\tau)d\tau$, that is, $\alpha(t)=2S_\gamma(t)$ in the notation of the main text.
The first-order errors are given by:
\begin{equation}
    \begin{split}
    A_1^{\delta H_J}(t)&=\delta J\left(A_1^{ZZ}(t)+A_1^{IZ}(t)\right)\\
A_1^{\delta H_\omega}(t) &= \delta \omega A_1^{IZ}(t)\\
    \end{split}
\end{equation}
where the contributions $A_1^{ZZ}(t)$ and $A_1^{IZ}(t)$ are:
\begin{equation}
\begin{split}
    A_1^{ZZ}(t)&=\int_0^t d\tau \Bigl[R_{\frac{IX-ZX}{2}}(-\alpha)R_{IX}(-\phi)R_{\frac{IY+ZY}{2}}(-\chi)R_{IX}(-\theta)\\
    &\qquad\qquad {}\times ZZ R_{IX}(\theta)R_{\frac{IY+ZY}{2}}(\chi)R_{IX}(\phi)R_{\frac{IX-ZX}{2}}(\alpha)\Bigr]\\
    &=ZZ\int_0^t d\tau \Bigl[R_{\frac{IX-ZX}{2}}(2\alpha)R_{IX}(\phi)R_{\frac{IY+ZY}{2}}(\chi)\\
    &\qquad\qquad {}\times R_{IX}(2\theta)R_{\frac{IY+ZY}{2}}(\chi)R_{IX}(\phi)\Bigr]\\
    A_1^{IZ}(t)&=\int_0^t d\tau \Bigl[R_{\frac{IX-ZX}{2}}(-\alpha)R_{IX}(-\phi)R_{\frac{IY+ZY}{2}}(-\chi)R_{IX}(-\theta)\\
    &\qquad\qquad {}\times IZ R_{IX}(\theta)R_{\frac{IY+ZY}{2}}(\chi)R_{IX}(\phi)R_{\frac{IX-ZX}{2}}(\alpha)\Bigr]\\
    &=IZ\int_0^t d\tau \Bigl[R_{\frac{IX-ZX}{2}}(2\alpha)R_{IX}(\phi)R_{\frac{IY+ZY}{2}}(\chi)\\
    &\qquad\qquad {}\times R_{IX}(2\theta)R_{\frac{IY+ZY}{2}}(\chi)R_{IX}(\phi)\Bigr]
\end{split}
\end{equation}
From here we can see that the integrals inside $A_1^{ZZ}(t)$ and $A_1^{IZ}(t)$ are identical, indicating that suppression of the $\delta\omega$-induced error $A_1^{\delta H_\omega}(T)$ also suppresses the $\delta J$ error $A_1^{\delta H_J}(T)$, but the reverse does not hold.

It should be noted that, since $IZ+ZZ$ only contributes to the upper-left block of the matrix, while $IX-ZX$ has nonzero elements only in the lower-right block, the term $R_{\frac{IX-ZX}{2}}(\alpha)$ does not affect the noise component $\delta J (IZ+ZZ)$. As a result, $A_1^{\delta H_J}(t)$ simplifies to:
\begin{equation}
    A_1^{\delta H_J}(t) =\delta J (IZ+ZZ) \int_0^t d\tau \left(R_{IX}(\phi)R_{\frac{IY+ZY}{2}}(\chi)R_{IX}(2\theta)R_{\frac{IY+ZY}{2}}(\chi)R_{IX}(\phi)\right)
\end{equation}
where the upper-left block of the integral is identical to the expression in Eq.~\eqref{eq:errorcase1}.

\subsection{Noise in Three-Qubit Chain}

For a three-qubit system, while a complete description of errors due to noise can become cumbersome, the error susceptibility exhibits a structure similar to the two-qubit case. Specifically, eliminating the error component of $A_1^{\delta H_\omega}$ also eliminates the error contributed by the coupling noise, $\delta J$. To illustrate this, consider the noisy Hamiltonian $H=H_0+\delta\omega IIZ$, expressed in block-diagonal form:
\begin{equation}
    H = \text{Diag}\{\tilde{H}_1, \tilde{H}_2,\tilde{H}_3,\tilde{H}_4\}+\delta\omega\text{Diag}\{Z,Z,Z,Z\}
\end{equation}
where the sub-blocks, omitting coefficients, take form of $\tilde{H}_{1/4}=\pm\beta Z + \Omega(t)X$ and $\tilde{H}_2=\tilde{H}_3=\Omega(t)X$. The symmetrical structure of the evolution for $\tilde{H}_{1}$ and $\tilde{H}_{4}$ ensures that these blocks share the same susceptibility to noise. To achieve robustness against noise, a control scheme must make all blocks resistant to errors.

For coupling noise, the noise Hamiltonian follows a similar block-diagonal structure, where each block receives a noise term of the form $\pm Z\delta J$, mirroring the form of frequency noise. Hence, as in the two-qubit case, any control pulse that suppresses the $\delta\omega$-induced error will also suppress the $\delta J$-induced error.

\subsection{Error Cancellation Condition}

We now derive an explicit condition for canceling first-order errors within a generic $2\times 2$ sub-block. As shown in the previous subsections, both $\delta H_J$ and $\delta H_\omega$ reduce, on each diagonal block, to an effective $\pm Z$ perturbation; in what follows we therefore work with the block-level operator $A_1=\int_0^T U_0^\dagger Z U_0\, dt$, with the understanding that the full multi-qubit susceptibility $A_1^{\delta H_k}$ defined in the main text is recovered by attaching the appropriate $ZZ$ or $IZ$ prefactor to each block.
For blocks with $\beta\neq0$, the evolution operator is:
\begingroup
\small
\begin{equation}
    U_0(\theta,\chi,\phi)=\left(
\begin{array}{cc}
 \cos \left(\frac{\chi }{2}\right) \cos \left(\frac{\theta +\phi }{2}\right)-i \sin \left(\frac{\chi }{2}\right) \sin \left(\frac{\theta -\phi }{2}\right) & -\sin \left(\frac{\chi }{2}\right) \cos \left(\frac{\theta -\phi }{2}\right)-i \cos \left(\frac{\chi }{2}\right) \sin \left(\frac{\theta +\phi }{2}\right) \\
 \sin \left(\frac{\chi }{2}\right) \cos \left(\frac{\theta -\phi }{2}\right)-i \cos \left(\frac{\chi }{2}\right) \sin \left(\frac{\theta +\phi }{2}\right) & \cos \left(\frac{\chi }{2}\right) \cos \left(\frac{\theta +\phi }{2}\right)+i \sin \left(\frac{\chi }{2}\right) \sin \left(\frac{\theta -\phi }{2}\right) \\
\end{array}
\right)
\end{equation}
\endgroup

Since both frequency and coupling noise reduce to a $Z$ component on this block (as noted above), the accumulated first-order error is
\begin{equation}
    A_1=\int_0^T U_0^\dagger(\theta(t),\chi(t),\phi(t)) Z U_0(\theta(t),\chi(t),\phi(t))\, dt.
\end{equation}
Decomposing it in the Pauli basis, and treating $\theta$, $\phi$, and time $t$ as functions of $\chi$, we obtain
\begin{equation}\begin{split}
A_X &= \int_0^{\chi_T} \left(-\cos(\theta)\sin(\chi)\right)t'(\chi)d\chi\\
A_Y &= \int_0^{\chi_T} \left(\sin(\theta)\cos(\phi) - \cos(\theta)\cos(\chi)\sin(\phi)\right)t'(\chi)d\chi\\
A_Z &= \int_0^{\chi_T} \left(\cos(\chi)\cos(\theta)\cos(\phi) - \sin(\theta)\sin(\phi)\right)t'(\chi)d\chi
\end{split}
\end{equation}

where $\theta(\chi)=\text{Arg}(\sin(\chi)\frac{d\phi}{d\chi}-i)+\pi$ follows from the $\Omega_y=0$ constraint, $t'(\chi)=\frac{1}{|\beta|}\sqrt{1+\sin(\chi)^2\left(\frac{d\phi}{d\chi}\right)^2}$ is the speed along the curve, and $\chi_T=\chi(T)$. Throughout this appendix we nondimensionalize time by $1/|\beta|$, where $\beta$ is the detuning of the nonzero block, and set $|\beta|=1$; the prefactor in $t'(\chi)$ is suppressed accordingly in what follows.

For the $\beta=0$ case, the first-order error in Magnus expansion is given by
\begin{equation}
    A_1 = \int_0^T R_X^\dagger(\Phi(\chi)) Z R_X(\Phi(\chi))t'(\chi)d\chi
\end{equation}
where $\Phi(\chi)=\Delta\theta + \int_0^{\chi(T)} \cos(\chi) \phi'(\chi) d\chi$. The Pauli components of the first-order error are therefore:
\begin{equation}
\begin{split}
    A_Z &= \int_0^{\chi(T)} \cos(\Delta\theta+\Delta\phi-2S_\gamma(\chi)) t'(\chi)d\chi\\
    A_Y &= \int_0^{\chi(T)} \sin(\Delta\theta+\Delta\phi-2S_\gamma(\chi)) t'(\chi)d\chi\\
    S_\gamma(\chi_t)&=\frac{1}{2}\int_0^{\chi_t} (1-\cos(\chi))\phi'(\chi)d\chi
\end{split}
\end{equation}

where $\Delta \theta=\theta(T)-\theta(0)$ and $\Delta\phi=\phi(T)-\phi(0)$.
We shall also consider the quantum crosstalk between two sub-blocks with $\beta>0$ and $\beta=0$. These two sub-blocks are coupled through the crosstalk term $\tilde{V}_{\text{cr}}\propto \Omega(t) (\cos(\tilde{\Delta} t)XZ+\sin(\tilde{\Delta}t) YZ)$; the amount of quantum crosstalk has the form
\begin{equation}
    A_1=\int_0^{\chi_T} \left(\frac{d\theta}{d\chi} + \cos(\chi)\frac{d\phi}{d\chi}\right) e^{i \tilde{\Delta} t(\chi)}U_0^\dagger(\theta,\chi,\phi) Z R_X(\Phi(\chi)) t'(\chi) d\chi
\end{equation}

It has components written in the form
\begin{equation}
    \begin{split}
        &\int_0^{\chi_T}  \left(\frac{d\theta}{d\chi} + \cos(\chi)\frac{d\phi}{d\chi}\right) \cos(\frac{\chi}{2})e^{-i(S_\gamma(\chi)-\theta-\phi)} e^{i \tilde{\Delta} t(\chi)} d\chi\\
&\int_0^{\chi_T}  \left(\frac{d\theta}{d\chi} + \cos(\chi)\frac{d\phi}{d\chi}\right)  \sin(\frac{\chi}{2})e^{i(S_\gamma(\chi)-\theta)} e^{i \tilde{\Delta} t(\chi)}d\chi\\
    \end{split}
\end{equation}

\appendixsection{Numerical Optimal Control}
\label{app:Parameters of Pulse}

Finding a function $\phi(\chi)$ to suppress all error terms caused by noise and crosstalk in a straightforward manner is generally challenging. However, using their analytical expressions, gradient-based numerical methods can efficiently minimize these terms. The objective function $C_\text{total}=w_1|C_\text{target}|^2+w_2|C_\text{robust}|^2$, the boundary conditions on $\phi(\chi;\vec{b})$, the explicit ansatz, and the optimized parameters are all given in the Methods section of the main text and are not repeated here. This appendix records the numerical implementation.

We recall only that $C_\text{target}=0$ is required for the three-qubit chain, and for a two-qubit system when the drive is resonant with one of the split frequencies; for two-qubit systems driven at the midpoint of the split frequencies it is satisfied by construction and can be omitted. The noise index runs over $k\in\{J,\omega,\epsilon\}$, corresponding to always-on coupling strength noise, qubit frequency noise, and coherent control crosstalk. The coherent control crosstalk Hamiltonian $\delta H_{\text{cr}}=\tilde{V}_\text{cr}$ is given in Eq.~\eqref{eq:crosstalktwo} or Eq.~\eqref{eq:crosstalkthree} for the two- and three-qubit cases, with $\epsilon$ the crosstalk strength.

\subsection{Numerical implementation}

The susceptibility integrals $A_1^{\delta H_k}(T)$ are evaluated in the curve parameter $\chi$ rather than in time, using the expressions of Appendix~\ref{app:Analysis of error curve}, so that the objective is a closed-form function of $\vec{b}$ and its gradient is obtained by automatic differentiation. The optimization is implemented in JAX~\cite{jax2018github} with Optax~\cite{deepmind2020jax}.

All curve integrals are evaluated on a uniform grid of $10^4$ points in $\chi$ over $[0,4\pi]$, with $\phi'(\chi)$ obtained by automatic differentiation of the ansatz rather than by finite differences. The parameters are optimized with Adam, applied as an Optax chain of gradient clipping by global norm $1$, Adam scaling ($\beta_1=0.9$, $\beta_2=0.999$, $\epsilon=10^{-8}$), and a warmup--cosine-decay learning-rate schedule that rises from zero to a peak of $10^{-1}$ over $200$ warmup steps and then decays quadratically to $10^{-3}$ over the remaining iterations. Runs use $10^3$ iterations with early stopping after $300$ iterations without improvement, random initialization, and retention of the best restart. In the composite objective the enclosed-solid-angle term carries weight $10^5$ and the three susceptibility terms (nonzero-$\beta$, zero-$\beta$, and control crosstalk) carry equal weight $10^3$, with an additional smoothness/regularization term of unit weight; the crosstalk term is evaluated at $\Delta=10$ in units of the coupling.

Fidelities are computed by piecewise-constant propagation on a uniform grid of $10^4$ time steps: the Hamiltonian is averaged over each interval and exponentiated exactly, and the resulting step propagators are composed in sequence. The two-dimensional noise maps scan a $30\times30$ grid over $|\delta\omega|,|\delta J|\le0.1$ in units of the coupling.

Two points about the reported numbers are worth stating explicitly. First, the pulse design uses the block-diagonal model of Eq.~(\ref{eq:qubitblock}), but every fidelity quoted in the main text is obtained by numerically propagating the full Hamiltonian of Eq.~(\ref{eq:heisenberg}), including the transverse coupling and the control crosstalk that the block model omits. The design model is therefore never used to score itself. Second, the waveforms plotted in the main-text figures are the physical drive $\Omega(t)$ rather than the dressed amplitude $\tilde{\Omega}(t)=\cos(\theta)\Omega(t)$; the two differ at $O(g^2/\Delta^2)$, as noted in Appendix~\ref{app:Simplification of the Hamiltonian}.

For scenarios where noise robustness is not required, setting all parameters to zero yields the simplest pulse; the shortest pulse satisfying $C_\text{target}=0$ is obtained by setting $b_2=c=0$ and $b_1=\frac{3465}{256}\pi(3+4\pi^2)a$, for which the constraint of the Methods section returns $b_3=0$.

\subsection{Dependence on transverse mixing}
To quantify the limitation imposed by residual inter-block control crosstalk, we evaluate the two-qubit robust pulse over a range of transverse mixing strengths $g/\Delta$, without reoptimizing the waveform. Figure C.1 compares the total gate infidelity with the error associated with $\tilde V_{\mathrm{cr}}$, estimated from the difference between the propagators with and without this term. For $g/\Delta\lesssim 0.05$, the residual crosstalk contribution is strongly suppressed. As $g/\Delta$ increases, this contribution becomes dominant, illustrating the limitation of the effective block-diagonal description.
\begin{figure}[h]
\centering
\includegraphics[width=0.65\textwidth]{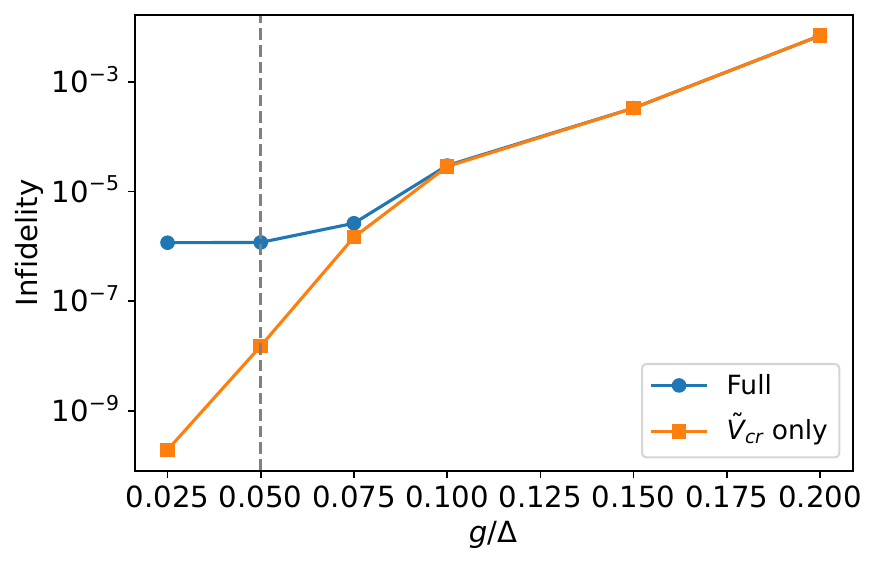}
\caption{\textbf{Effect of transverse inter-qubit mixing on gate fidelity.} Total gate infidelity and residual inter-block crosstalk error as functions of $g/\Delta$ for the two-qubit robust pulse. The latter is quantified by $1-F_{\mathrm{avg}}(U_{\mathrm{full}},U_{\mathrm{block}})$, where $U_{\mathrm{full}}$ includes $\tilde V_{\mathrm{cr}}$ and $U_{\mathrm{block}}$ omits it. The pulse is not reoptimized as $g/\Delta$ varies. The vertical dashed line indicates the parameter $g/\Delta=1/20$ used in the main numerical demonstrations.}\label{fig:transversecoupling}
\end{figure}

\appendixsection{Generalized Windings and Entangling Gates}
\label{app:different windings}

\subsection{General Winding Ansatz}
A geometric curve forming a loop on the 2-sphere can wrap around it multiple times. In the main text (and previous section), we adopted a control pulse whose curve completes two full revolutions ($\chi \in [0, 4\pi]$). Here, we provide a more general ansatz that accommodates an arbitrary number of windings, producing pulse shapes with varied lengths and amplitudes that may be more adaptable to diverse experimental conditions. This ansatz also serves as a tool for constructing multi-qubit entangling gates.

In a general setting, $\phi(\chi)$ is defined on the interval $[0, \chi_T]$, where $\chi_T=2m\pi$ and $m$ denotes the number of times the curve wraps around the sphere in the polar direction. The ansatz is composed of polynomial and Fourier terms:
\begin{equation}
\begin{split}
    \phi(\chi)&=\Phi \left(\frac{\chi}{\chi_T}\right)^2\left(3-2\frac{\chi}{\chi_T}\right) \\ &+ \sum_{i=1}^n\left(a_i \cos\left((2i-1)\pi\frac{\chi}{\chi_T}\right)+b_i \sin\left(2i\pi\frac{\chi}{\chi_T}\right) \right)
    \end{split}
\end{equation}
Here $n$ is the number of Fourier components controlling how the curve encircles the sphere in the azimuthal direction, and the parameters $a_i, b_i$ are tuned to optimize pulse characteristics. To ensure the pulse begins and ends at zero, we impose
\begin{equation}
    \sum_{i=1}^n a_i = 0,\quad
    \sum_{i=1}^n i b_i = 0.
\end{equation}
Additionally, we can constrain the solid angle subtended by the curve via $\int_0^{\chi_T}(1-\cos(\chi))\phi'(\chi)\, d\chi=2S_\gamma$, where $S_\gamma$ is the half signed solid angle defined in Eq.~(\ref{eq:zeroarea}) of the main text. This yields:
\begin{equation}
    2 \chi_T^2 \left(\sum _{i=1}^n \frac{a_i}{\pi ^2 (2 i-1)^2-\chi_T^2}\right)-\pi  \sum _{i=1}^n i b_i \delta (\chi_T-2 i \pi )+\Phi +\frac{12 \Phi }{\chi_T^2}=2S_\gamma
\end{equation}

where $\delta(x)$ is the discrete delta function, equal to 1 when $x=0$ and 0 otherwise. The constraint then simplifies to
\begin{equation}
    2S_\gamma =
    \begin{cases}
        2 \chi_T^2 \left(\sum _{i=1}^n \frac{a_i}{\pi ^2 (2 i-1)^2-\chi_T^2}\right)-\pi m b_m+\Phi +\frac{12 \Phi }{\chi_T^2} & n\geq m \\
        2 \chi_T^2 \left(\sum _{i=1}^n \frac{a_i}{\pi ^2 (2 i-1)^2-\chi_T^2}\right)+\Phi +\frac{12 \Phi }{\chi_T^2} & n < m \\
    \end{cases}
\end{equation}
These constraints leave $n-2$ coefficients $a_i$ and $n-1$ coefficients $b_i$ as free parameters, while the remaining coefficients ($a_{n-1}$, $a_{n}$, and $b_n$) are determined by solving the linear system above.

\begin{figure}[h]
    \centering
    \includegraphics[width=0.9\linewidth]{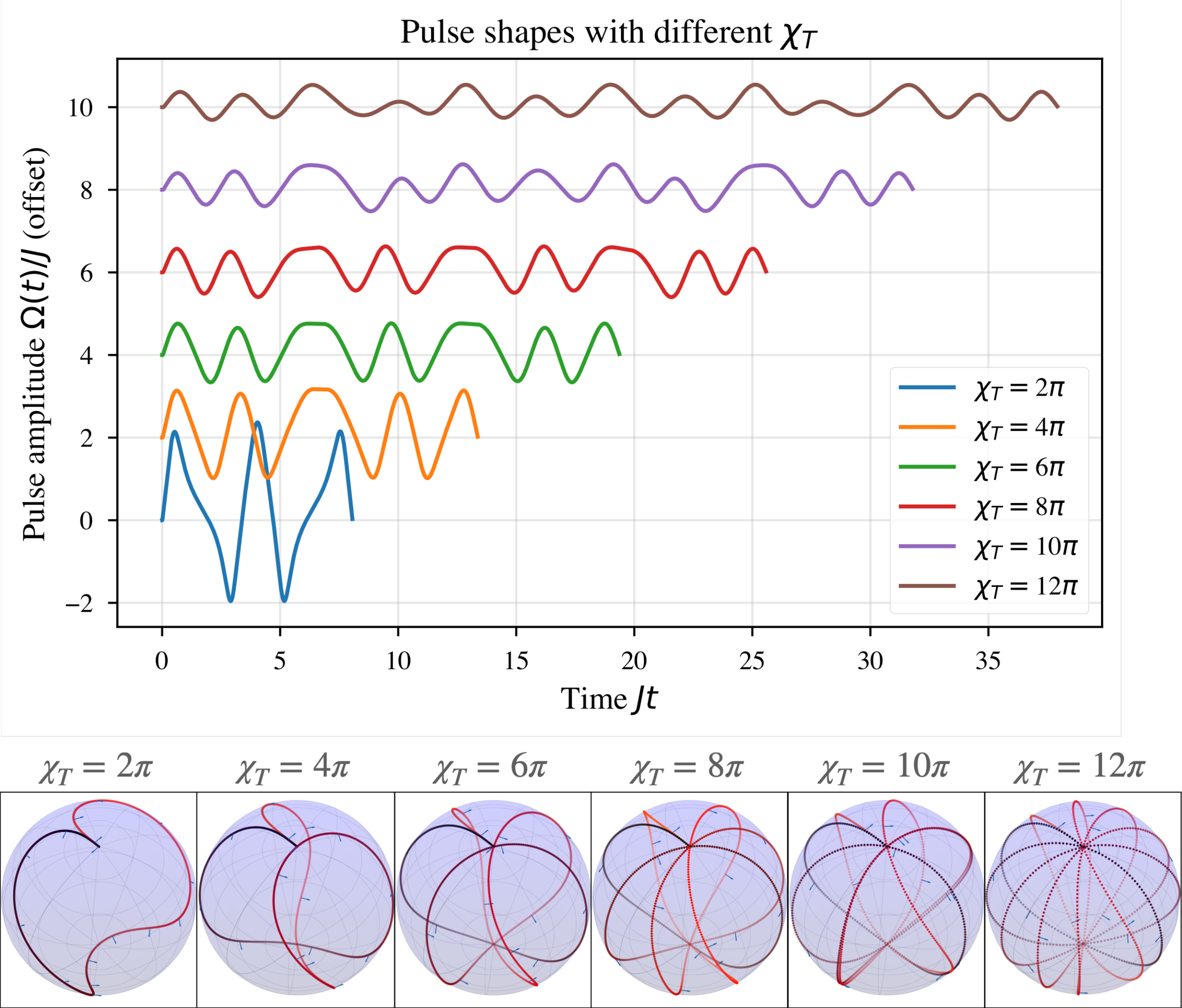}
    \caption{\textbf{Pulse waveforms for different winding numbers.} Pulse waveforms for single-qubit $\pi$-gates generated using different winding numbers ($m=1$ to $6$). Curves are vertically offset by increments of 2 units for clarity.}
    \label{fig:different_chi_t}
\end{figure}

We demonstrate the generalized approach by showing pulse waveforms and the corresponding geometric curves for single-qubit $X(\pi)$ gates (with $S_\gamma=0$), for various winding numbers $m = 1$ to $6$ (i.e., $\chi_T=2m\pi$), as illustrated in Fig.~\ref{fig:different_chi_t}.

\subsection{Entangling-Gate Construction}

The same geometric construction produces entangling gates once the closed loop is allowed to subtend a nonzero signed solid angle. As noted in Sec.~\ref{sec:geometry}, a loop with $S_\gamma\neq0$ leaves the detuned blocks with rotation angle $\Phi$ and the resonant blocks with $\Phi-2S_\gamma$. Writing $R_{ZX}(\vartheta)\equiv\exp(-i\vartheta\, Z\otimes X/2)$, with $Z$ acting on the spectator qubit and the orientation fixed so that the detuned block is the $Z=+1$ sector (the block labelling of Appendix~\ref{app:Simplification of the Hamiltonian}), the two-qubit evolution is
\begin{equation}\label{eq:entangler}
    U = R_{IX}(\Phi-S_\gamma)\,R_{ZX}(S_\gamma).
\end{equation}
The nonlocal content of the gate is therefore set by $S_\gamma$ alone, and is maximal at $S_\gamma=\pi/2$. Note that $S_\gamma=\pi$ gives $R_{ZX}(\pi)=-i\,Z\otimes X$, a product of Pauli operators, which is not entangling.

Choosing $\Phi=\pi$ together with $S_\gamma=\pi/2$ makes the detuned block a $\pi$-rotation and the resonant block the identity, so that
\begin{equation}\label{eq:cnot}
    U_{2q}=e^{-i\pi/4}\left[R_Z(\pi/2)\otimes I\right]\mathrm{CNOT},
\end{equation}
an exact CNOT up to a global phase and a $Z$ rotation on the spectator qubit, the latter implemented virtually in software at no cost in gate time.

For a three-qubit chain the same pulse acts on the four blocks $\beta=\{J,0,0,-J\}$, flipping the central qubit when its two neighbors are in the same logical state ($|00\rangle$ or $|11\rangle$) and leaving it unchanged otherwise; we term this an XNOR-CNOT gate. Here the phase of the $\pi$-rotation is common to both same-state sectors, giving
\begin{equation}\label{eq:xnorcnot}
    U_{3q}=e^{-i\pi/4}\,e^{-i\frac{\pi}{4}Z_1Z_2}\;U_{\text{XNOR-CNOT}},
\end{equation}
where the residual factor is a controlled-$Z$ gate between the two spectators, up to single-qubit $Z$ rotations. Unlike the two-qubit case this correction is not removable by local operations; it is a fixed Clifford operation that can be absorbed at the circuit level, and we state it explicitly so that the realized unitary is unambiguous.

\begin{figure}[h]
\centering
\includegraphics[width=0.9\textwidth]{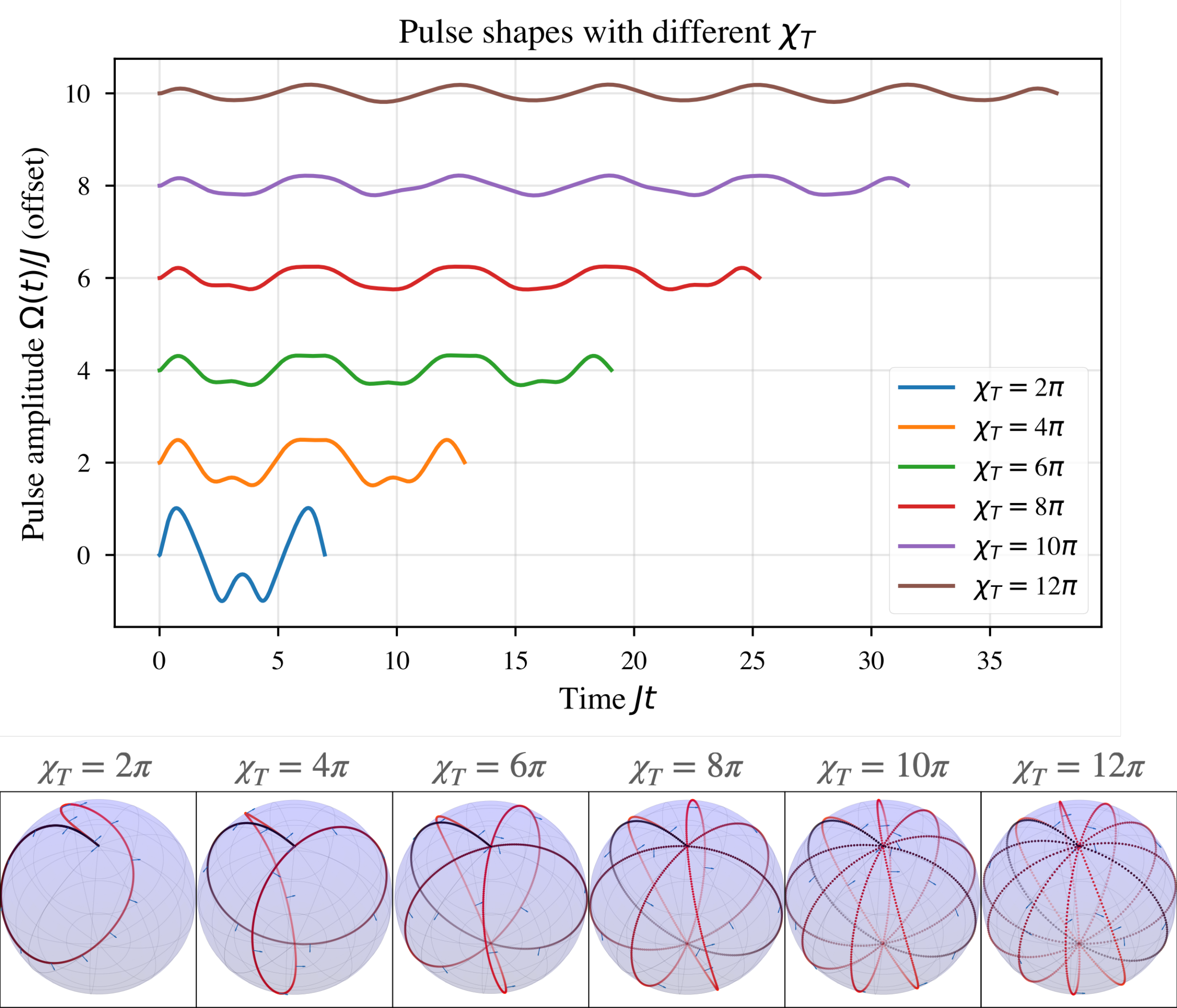}
\caption{\textbf{Entangling gates from nonzero enclosed solid angle.} Pulse waveforms (top) and the corresponding 2-sphere trajectories (bottom) for the entangling construction with $\Phi=\pi$ and $S_\gamma=\pi/2$, shown for the six winding numbers $\chi_T=2m\pi$ with $m=1$--$6$. Each waveform realizes an exact CNOT in the two-qubit system and the XNOR-CNOT gate in the three-qubit chain, up to the local and $Z_1Z_2$ factors of Eqs.~(\ref{eq:cnot}) and (\ref{eq:xnorcnot}). Waveforms are vertically offset by increments of 2 units for clarity; trajectories are color-coded from red at the start to black at the end.}\label{fig:entangling}
\end{figure}

Figure~\ref{fig:entangling} shows the resulting pulse waveforms and their 2-sphere trajectories. For spectator-independent single-qubit gates, the enclosed solid angle must vanish when resonant subspaces are present. Allowing a nonzero solid angle instead produces conditional rotations that can generate entangling operations. These constructions realize the target operations within the effective block-diagonal model; their noise robustness is not analyzed here.

\appendixsection{Perturbative Robust Control Pulse Benchmark}
\label{app:pRCP}

The perturbative robust control pulse (pRCP) used as a benchmark in the main text is representative of the family of analytical noise-robust control schemes that treat the always-on inter-qubit coupling as a small perturbation around an isolated qubit. In these constructions, the noise-free dynamics is approximated as that of an isolated qubit subject to a transverse drive, and the pulse waveform is shaped so that the first-order Magnus-expansion error vector in $\mathbb{R}^3$ closes on itself, cancelling the linear sensitivity to a transverse noise channel. The perturbative Euclidean error-curve condition follows from the small-$\beta$ expansion of the exact gate-matching condition, which at first order requires $\int_0^T U_0^\dagger(t)ZU_0(t) dt=0$.

We adopt the explicit pulse ansatz constructed in Ref.~\cite{hai2025geometric} for a $\pi$-gate:
\begin{equation}\label{eq:prcp}
\Omega_\text{pRCP}(t) = \frac{1}{T}\left[a_0 + a_2\cos\!\left(\frac{2\pi t}{T}\right) + a_4\cos\!\left(\frac{4\pi t}{T}\right)\right]\sin\!\left(\frac{\pi t}{T}\right),
\end{equation}
with coefficients
\begin{equation}
(a_0, a_2, a_4) \approx (0.5108,\, -12.94,\, -1.6595),
\end{equation}
and the gate time $T$ chosen so that the maximum amplitude $\max_t |\Omega_\text{pRCP}(t)|$ matches that of our robust pulse. The resource held fixed in the comparison is therefore peak drive power, with gate duration allowed to float; the coefficients $(a_0,a_2,a_4)$ are those of Ref.~\cite{hai2025geometric} and are not reoptimized here, since they are fixed by the first-order cancellation conditions of that construction and only the overall time scale is free. The overall $\sin(\pi t/T)$ envelope enforces smooth onset and shutoff ($\Omega_\text{pRCP}(0) = \Omega_\text{pRCP}(T) = 0$), while the two cosine harmonics implement the first-order noise-cancellation conditions of Ref.~\cite{hai2025geometric}.

This pulse is by construction first-order robust against weak transverse noise ($\delta\omega Z$ in our notation), where the ZZ-induced detuning $\beta$ acts as a small parameter that can be Taylor-expanded around zero. In the regime used in the main numerical comparisons, however, the coupling-induced detuning $\beta \sim J$ is no longer small compared to the drive bandwidth. The pRCP then leaves a substantial systematic infidelity floor visible in Figs.~\ref{fig:PiPulse_2} and \ref{fig:PiPulse_3}, even though it remains designed to suppress transverse noise to first order. Our robust pulse targets the same noise channels but does so via the full spherical conditions. The fidelity gap between the two quantifies the gain from a non-perturbative geometric description.

\bibliographystyle{unsrt}
\bibliography{Reference}

\end{document}